\documentclass[]{aa}
\usepackage[utf8]{inputenc}
\usepackage{natbib}
\usepackage{hyperref}
\usepackage{graphicx}
\usepackage{lscape}
\usepackage{pdflscape}
\usepackage{txfonts}
\usepackage{placeins}

\title{Multiplicity of northern bright O-type stars with optical long baseline interferometry}
\subtitle{Results of the pilot survey}
\titlerunning{Multiplicity of northern bright O-type stars with OLBI}

\author{C.~Lanthermann\inst{1,2,3}
        \and
        J.-B.~Le Bouquin\inst{3}
        \and
        H.~Sana\inst{2}
        \and
        A.~Mérand\inst{4}
        \and
        J. D.~Monnier\inst{5}
        \and
        K.~Perraut\inst{3}
        \and
        A.J.~Frost\inst{2}
        \and
        L.~Mahy\inst{6}
        \and
        E.~Gosset\inst{7}
        \and
        M.~De~Becker\inst{7}
        \and
        S.~Kraus\inst{8}
        \and
        N.~Anugu\inst{1}
        \and
        C. L. Davies\inst{8}
        \and
        J. Ennis\inst{5}
        \and
        T. Gardner\inst{5}
        \and
        A. Labdon\inst{9}
        \and
        B. Setterholm\inst{5}
        \and
        T. ten Brummelaar\inst{1}
        \and
        G. H. Schaefer\inst{1}
    }

\institute{The CHARA Array of Georgia State University,         Mount Wilson Observatory, Mount Wilson, CA              91023, USA
    \and
        Institute of Astronomy, KU Leuven, Celestijnenlaan 200D, 3001 Leuven, Belgium
    \and
        Institut de Planétologie et d'Astrophysique de Grenoble, Grenoble 38058, France
     \and
        European Southern Observatory, Karl-Schwarzschild-Str. 2, 85748 Garching, Germany
     \and
         University of Michigan, Department of Astronomy, 1085 S. University Ave; West Hall 323; Ann Arbor, MI 48109
     \and
         Royal Observatory of Belgium, Avenue Circulaire/Ringlaan 3, B-1180, Brussels, Belgium
     \and
         Space sciences, Technologies, and Astrophysics Research (STAR) Institute, University of Liège, Quartier Agora, 19c, Allée du 6 Août, B5c, B-4000 Sart Tilman, Belgium
     \and
         Astrophysics Group, School of Physics and Astronomy, University of Exeter, Stocker Road, Exeter, EX4 4QL, UK
     \and
         European Southern Observatory, Casilla 19001, Santiago 19, Chile
\\ correspondence: \texttt{clanthermann@gsu.edu}
   }
   
\date{November 2022}

\abstract{
The study of the multiplicity of massive stars gives hints on their formation processes and their evolutionary paths, which are still not fully understood. Large separation binaries (>50 milliseconds of arc, mas) can be probed by adaptive-optics-assisted direct imaging and sparse aperture masking, while close binaries can be resolved by photometry and spectroscopy. However, optical long baseline interferometry is mandatory to establish the multiplicity of Galactic massive stars at the separation gap between 1 and 50 mas.
}{
In this paper, we aim to demonstrate the capability of the new interferometric instrument MIRC-X, located at the CHARA Array, to study the multiplicity of O-type stars and therefore probe the full range of separation for more than 120 massive stars ($H<7.5$~mag).
}{
We initiated a pilot survey of bright O-type stars ($H<6.5$~mag) observable with MIRC-X. We observed 29 O-type stars, including two systems in average atmospheric conditions around a magnitude of $H=7.5$~mag. We systematically reduced the obtained data with the public reduction pipeline of the instrument. We analyzed the reduced data using the dedicated python software CANDID to detect companions.

}{
Out of these 29 systems, we resolved 19 companions in 17 different systems with angular separations between $\sim0.5$ and 50~mas. This results in a multiplicity fraction $f_{\text{m}}=17/29=0.59\pm0.09$, and an average number of companions $f_{\text{c}}=19/29=0.66\pm0.13$. Those results are in agreement with the results of the SMASH+ survey in the Southern Hemisphere. Thirteen of these companions have been resolved for the first time, including the companion responsible for the nonthermal emission in Cyg~OB2-5~A and the confirmation of the candidate companion of HD~47129 suggested by SMASH+.

}{
A large survey on more than 120 northern O-type stars ($H<7.5$) is possible with MIRC-X and will be fruitful.
}

\begin{document}           

\maketitle

\section{Introduction}

Massive stars are key components of the evolution of their host galaxy. They are the main producers of heavy elements, and the momentum and kinetic energy  involved in their death have an influence on a large part of their galaxy~\citep{2007ARA&A..45..481Z}. They are also the progenitors of the compact objects that, when they merge, produce gravitational wave bursts that we can currently detect~\citep{2016PhRvL.116f1102A}.

However, their short lifetime (a few million years) and their rapid formation process ($10^5$ years) make the observation of their early ages difficult~\citep{2014prpl.conf..149T}. Indeed, their lifetime makes them rare, and so to observe a large number of massive stars (> 100), one needs to look for them at significant distances, typically 1 to 3 kpc. In addition, the majority of young massive stars are still embedded in a cloud of gas and dust when they finish their formation process~\citep{zinnecker_2005}, making the observation of this formation step even harder. In consequence, the formation process of massive stars is still actively discussed.

Historically, the standard models of star formation could not explain the formation of stars with masses significantly higher than about 10 M$_{\sun}$. The main difficulty is overcoming the radiation barrier emitted by the protostar as soon as it starts burning nuclear fuel. So, specific formation models need to explain how massive stars can form. There are currently three main scenarios: 1) the core accretion~\citep{terquem_2001, 2002ASPC..267..165Y}, which uses a massive accretion disk to accrete more matter; 2) the competitive accretion~\citep{1978MNRAS.184...69L, 1982NYASA.395..226Z, 10.1046/j.1365-8711.2002.05794.x, 2003MNRAS.343..413B} for which close protostar cores use the combined gravitational potential to attract matter from further away than with each individual gravitational potential; and 3) the collision~\citep{1988Natur.333..219B,2005MNRAS.362..915B, 2006MNRAS.366.1424D} in which intermediate-mass stars collide to merge and form a more massive star. The outcomes of these three different formation models predict different multiplicity parameters. Hence, the study of the multiplicity of massive stars, after their formation process, should provide relevant constraints on these formation models. 

Another motivation for the investigation of the multiplicity of massive stars, especially in the range of periods considered in this study, is the investigation of physical processes driven by the colliding winds. In particular, such systems are well suited for particle acceleration, hence the class of particle-accelerating colliding-wind binaries (PACWBs, \citet{2013A&A...558A..28D,2017A&A...600A..47D}). Such systems, mainly revealed by synchrotron radio emission, are likely contributors to the population of lower energy Galactic cosmic rays. Appropriate knowledge of their orbit is required to interpret their behavior and model their shock physics.

To probe the full range of orbital separations of systems situated at a typical distance of 2 kpc, one needs to use different observational techniques. The close companions \citep[up to 0.5~milliseconds of arc, mas,][]{2020A&A...634A.119M} can be probed by photometry (eclipsing binaries) or spectroscopy \citep[radial velocity,][]{2012Msngr.148...33S,2013A&A...550A.107S,2017IAUS..329...89B}, while wide companions (separation > 50~mas) can be probed with techniques such as adaptive-optics-assisted direct imaging, aperture masking, speckle imaging, and coronography \citep{2008AJ....136..554T, 2009AJ....137.3358M, 2022A&A...660A.122R}. For separations between 1 and 50~mas, the only technique we can use is optical long baseline interferometry (OLBI) (see Fig.~1 in \citet{2017IAUS..329..110S}). But until recently, this technique was limited by its sensitivity and could only be applied to a modest sample of massive stars.

The advent of the PIONIER (Precision Integrated-Optics Near-infrared Imaging ExpeRiment) instrument at the very large telescope interferometer (VLTI)~\citep{2011A&A...535A..67L} enabled the southern massive stars at high angular resolution (SMASH+), the first systematic interferometric large survey on massive stars in the Southern Hemisphere \citep{Sana2014}, probing the missing range of separation to have a complete statistical study of the multiplicity of massive stars. With this survey, \citet{Sana2014} observed 96 southern O-type star systems, nearly reaching the 100 targets required to get a statistical error < 5\% over the entire range of the multiplicity fraction. The observational constraints brought by the SMASH+ survey, especially the abundance of companions with a separation smaller than 100~AU, are qualitatively in overall agreement with the core accretion model leading to disk fragmentation. However, the statistic on subgroups of stars, such as masses or evolutionary stage, is too low to obtain a robust conclusion. Therefore, we aim to perform a similarly large survey (120 objects) in the Northern Hemisphere to double the total statistic.

\begin{figure}[t]
\centering
\includegraphics[width=\columnwidth]{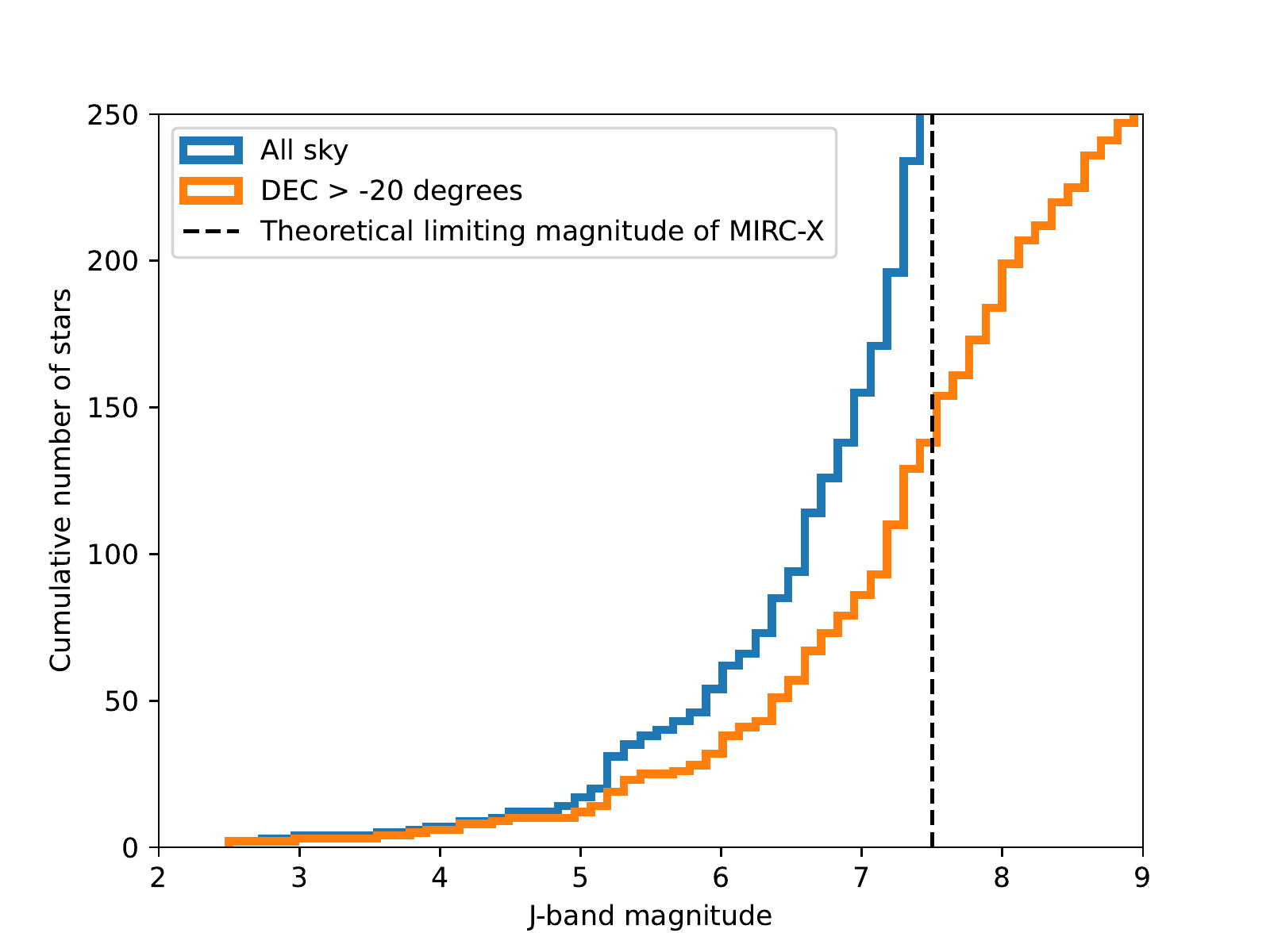}
\caption{Cumulative histogram of the number of O-type stars as a function of their magnitude in the J-band. In blue, for the whole sky. In orange, for a declination greater than $-20^\circ$ to be reachable with the CHARA Array. Data from the Galactic O-star catalog (GOSC)~\citep{GOSC}.} \label{fig:OvsHmag_cat}
\end{figure}

To observe more than 100 O-type stars in the Northern Hemisphere, one needs to reach a limiting magnitude in the J-band (the spectral band available in the GOSC catalog,~\citealt{GOSC}) of $J = 7.5$, as shown in Fig.~\ref{fig:OvsHmag_cat}. In this figure, the limit of declination > $-20^{\circ}$  corresponds to the limit of observability of the CHARA (center for high angular resolution astronomy) Array. Thanks to the recent implementation of the MIRC-X (Michigan InfraRed Combiner-eXeter) instrument~\citep{MIRCX,anugu2020,2019A&A...625A..38L} at the CHARA Array~\citep{CHARA}, located at Mount Wilson observatory, USA, this magnitude is now reachable in the H-band with the OLBI technique in the Northern Hemisphere. As O-type stars are hot stars and under typical reddening conditions, their magnitudes remain relatively the same in all the infrared spectral bands~\citep{2006A&A...457..637M}, meaning that the limit on the magnitude of 7.5 in the J-band required to observe >100 O-type objects remains valid in the H-band.

In this paper, we present the results of the pilot survey performed on 29 systems, with the goal to demonstrate the feasibility of a large survey of more than 100 O-type stars with the CHARA/MIRC-X instrument. We present the observations in Section~\ref{sec:Obs}, with the definition of the sample, the description of the observation campaign, and the data reduction process. We then describe the data analysis that we performed in Section~\ref{sec:dataanal}. Section~\ref{sec:Resdet} presents our results. We perform a statistical analysis of the results in Section~\ref{sec:statanal}. We finally discuss the results in Section~\ref{sec:discussion} and conclude in Section~\ref{sec:conclusion}.

\section{Observations}\label{sec:Obs}

\subsection{Observation sample}\label{subsec:Obs_sample}

We built our sample using the Galactic O-star catalog~\citep[GOSC][]{GOSC}. We selected every O-star with a declination DEC > $-20^\circ$ and with a magnitude in the J-band registered in GOSC of $J < 7.0$~mag. We chose the J-band criterion for several reasons. The first one is that the magnitude criteria available in GOSC are either in the B-band or the J-band. As we used the MIRC-X instrument, working in the H-band, the J-band is the nearest one of the two available. We adopted the threshold value of 7.0 or brighter because, during this pilot study, the official limiting magnitude offered by MIRC-X was $H = 6.5$~mag. As only the Rayleigh-Jeans tail of the O-type stars is observed in the J- and H-bands, their magnitudes are comparable in both bands. We took an extra 0.5 magnitude as a margin to be sure that our sample contains all targets observable with MIRC-X. We then looked for the H-band magnitude of the selected systems in the 2MASS All-Sky Catalog of Point Sources~\citep{2003yCat.2246....0C} and performed the last selection on stars with a magnitude $H < 6.5$ to comply with the MIRC-X limiting magnitude. Our input sample is therefore magnitude limited. We note that some of the targets in our sample overlap with targets already observed by SMASH+, which can be used to validate our results.

\subsection{Observation campaign}\label{subsec:obs_camp}

\begin{table*}[t!]
\caption{List of O-type stars observed with good quality data during the pilot survey.}
 \label{tab:logObs}
 \centering
 \begin{tabular}{lllllllll} 
 \hline \hline 
Name & Spectral type & $H$ & $d$ & SC & SIC & IC & WC & Total  \\
&  & (mag) & (kpc) & (<0.5 mas) & \multicolumn{2}{c}{(0.5 - 50 mas)}&(0.05\arcsec-8\arcsec)\\
\hline
Cyg OB2-5 A (BD+40 4220A) & O7Iafep & 4.745 & 1.748  &1 &0&2&2&5 \\
Cyg OB2-9 (HIP 101419) & O4.5If & 5.897 & 1.788 & 0&1&0&1&2 \\ 
Cyg OB2-10 (BD+41 3804) & O9.7Iab$^{\dagger}$ & 5.839 & 1.376 & 0&0&1&3&4\\
HD~17505 A & O6.5IIIn(f) & 6.177 & 2.258 & 1&0&1&1&3  \\ 
HD~19820 (CC Cas) & O8.5III(n)((f)) & 6.003 & 1.135 & 1&0&2&0&3  \\
HD~24431 & O9III & 5.845 & 0.922 & 0&0&1&1&2 \\ 
HD~28446 A (1 Cam A) & O9.7IIn & 5.459 & 0.760 & 1&0&1&1&3 \\ 
HD~30614 ($\alpha$ Cam) & O9Ia & 4.242 & 1.717 & 1&0&0&0&1 \\
HD~34078 (AE Aur) & O9.5V & 5.355 & 0.382 & 0&0&1&0&1 \\ 
HD~36861 ($\lambda$ Ori A) & O8III((f)) & 3.769 & 0.438 & 0&0&1&1&2 \\ 
HD~45314 & O9:npe & 5.761 & 0.864 & 0&0&1&0&1 \\
HD~47129 & O8I+O7.5III$^\ddagger$ & 5.806 & 1.283 & 1&0&1&2&4 \\ 
HD~47432 (V689 Mon) & O9.7Ib & 5.949 & 1.596 & 0&0&0&1&1 \\
HD~47839 (15 Mon AaAb) & O7V+B1.5/2V$^\star$ & 5.322 & 3.974 & 0&0&0&1&1  \\ 
HD~167971 (MY Ser AaAb) & O8Iaf(n)+O4/5 & 5.315 & 1.339 & 1&0&1&0&2  \\
HD~188001 (9 Sge) & O7.5Iabf & 6.166 & 1.833 & 0&0&0&0&0  \\
HD~193322 AaAb & O9IV(n) & 5.688 & 1.001 & 1&0&1&3&5 \\
HD~195592 & O9.7Ia & 4.911 & 1.729 & 1&0&0&0&1  \\ 
HD~201345 & ON9.2IV & 8.171 & 1.828 & 0&0&0&1&1 \\
HD~202214 AaAb & O9.5IV & 5.505 & 1.032 & 1&0&1&1&3 \\ 
HD~206183 & O9.5IV-V & 7.193 & 0.901 & 0&0&0&2&2   \\
HD~206267 AaAb & O6.5V((f))+O9/B0V & 5.254 & 0.789 & 1&0&0&4&5 \\
HD~207198 & O8.5II & 5.318 & 0.978 & 0&0&1&1&2 \\
HD~209975 (19 Cep) & O9Ib & 4.935 & 0.959 & 0&0&0&4&4 \\ 
HD~210809 & O9Iab & 7.401 & 3.661 & 0&0&0&0&0 \\
HD~210839 ($\lambda$ Cep) & O6.5I(n)fp & 4.618 & 0.832 & 0&0&0&0&0 \\
HD~217086 & O7Vnn((f))z & 6.100 & 0.830 & 0&0&0&2&2 \\ 
HD~228779 & O9Iab$^\dagger$ & 5.834 & 1.653 & 0&0&0&0&0 \\
HD~229196 & O6II(f) & 6.079 & 1.720 & 0&0&1&0&1 \\ 

\hline
\end{tabular}
\tablefoot{The first column provides the identifier of the star. The second column contains the spectral type, with the ones marked with a $\dagger$ coming from~\citet{2016ApJS..224....4M}, the one marked wit a $\ddagger$ coming from \citet{2011A&A...525A.101M}, and the one marked with a $\star$ coming from \citet{2013yCat....1.2023S}, the others coming from \citet{Sota_2011}. The third column displays the magnitude in the H-band as found in the 2MASS All-Sky Catalog of Point Sources~\citep{2003yCat.2246....0C}. The fourth column gives the distance that separates us from the system according to \citet{2021yCat.1352....0B}, except for HD~202214 for which the distance comes from \citet{2009A&A...507..833M}. The fifth to  eighth columns give the number of already known spectroscopic companions (SC), spectroscopic companions resolved by interferometry (SIC), interferometric companion (IC), and wide companion (WC), respectively. The references for already known companions can be found in the star-by-star description in Sect.~\ref{sec:Resdet}. The ninth column gives the total number of known companions after this study.}
\end{table*}

The MIRC-X beam combiner operates in the J- and H-band. The observations presented in this paper are performed only in the H-band (1.65 $\mu$m) because the J-band mode was still experimental during the pilot survey and only uses four telescopes. The six telescopes provide sufficient coverage of the uv-plane to constrain the multiplicity of a star in a single snapshot. This can also be done for data that combine only five telescopes when the conditions (weather, technical, operational) would not permit a six-telescope observation. The MIRC-X combiner allows different spectral resolutions. Our data have been taken with the PRISM-50 configuration, allowing a spectral resolving power of $R \sim 50$. This configuration was chosen to optimize the sensitivity of the beam combiner and because it brings the Outer Working Angle (OWA) of MIRC-X to: 
\begin{equation}\label{eq:OWA}
    OWA = R \frac{\lambda}{B} = 2.7 \times 10^{-7}~\text{rad} \simeq 55~\text{mas}
\end{equation}
where  $\lambda$ is the central wavelength which in the H-band is equal to $1.6 \times 10^{-6}$~m, and $B$ is the length of the baseline, equal to 330~m for the longest baseline at CHARA. This OWA allowed us to fill the gap in angular separation that other techniques cannot reach. We note that detection of companions with a separation larger than the OWA is still possible, but the flux ratio will be biased by the bandwidth smearing, hence, hampering accurate measurement of the contrast~\citep{2016SPIE.9907E..3BH}.

We note that the Inner Working Angle is the angular resolution of the instrument, meaning that a binary separated by less than this angle would not be resolved, and is defined by:
\begin{equation}\label{eq:IWA}
    IWA = \frac{\lambda}{2B} \simeq 0.55~\text{mas}.
\end{equation}
This angular resolution is about a factor of two to three smaller than the SMASH+ survey owing to the larger baseline $B$ of CHARA compared to the VLTI.

For the calibration strategy, we alternated a calibrator with a science target. The calibrator was chosen with the tool \textsc{searchcal}~\citep{searchcal}, developed by the JMMC\footnote{\href{https://www.jmmc.fr/english/tools/proposal-preparation/search-cal/}{https://www.jmmc.fr/english/tools/proposal-preparation/search-cal/}}. The selection criterion was the calibrators needed to be at most 1.5 magnitudes brighter and 0.5 magnitudes fainter than the science target it would calibrate, and situated at a maximum angular distance of 3 degrees on the sky. Most of our calibration stars are of spectral type KIII, for which a sufficiently accurate diameter (a few percent) can be estimated from the apparent photometry~\citep{2016A&A...589A.112C}.

The observations with MIRC-X have been carried out during five runs spread over three observation semesters from 2018 to 2019. The observation time was obtained through NOAO\footnote{\href{https://noirlab.edu}{https://noirlab.edu}} (now called NOIRLab) community access time (program IDs: 2018A-M12/NOAO2, 2018B-M17/NOAO4, 2019A-M16/NOAO4; PI: C. Lanthermann). We also used one night that studied O-type stars during which O-type stars were observed as backup targets because the original program could not be executed, in December 2017 (PI: S. Kraus).

During this campaign, we could obtain good quality data on 29 O-type stars, listed in Table~\ref{tab:logObs} with information on the spectral type, distance, and the number of detected companions in various separation ranges, from the literature as well as those detected by this study.
\begin{figure}[!t]
\centering
\includegraphics[width=\columnwidth]{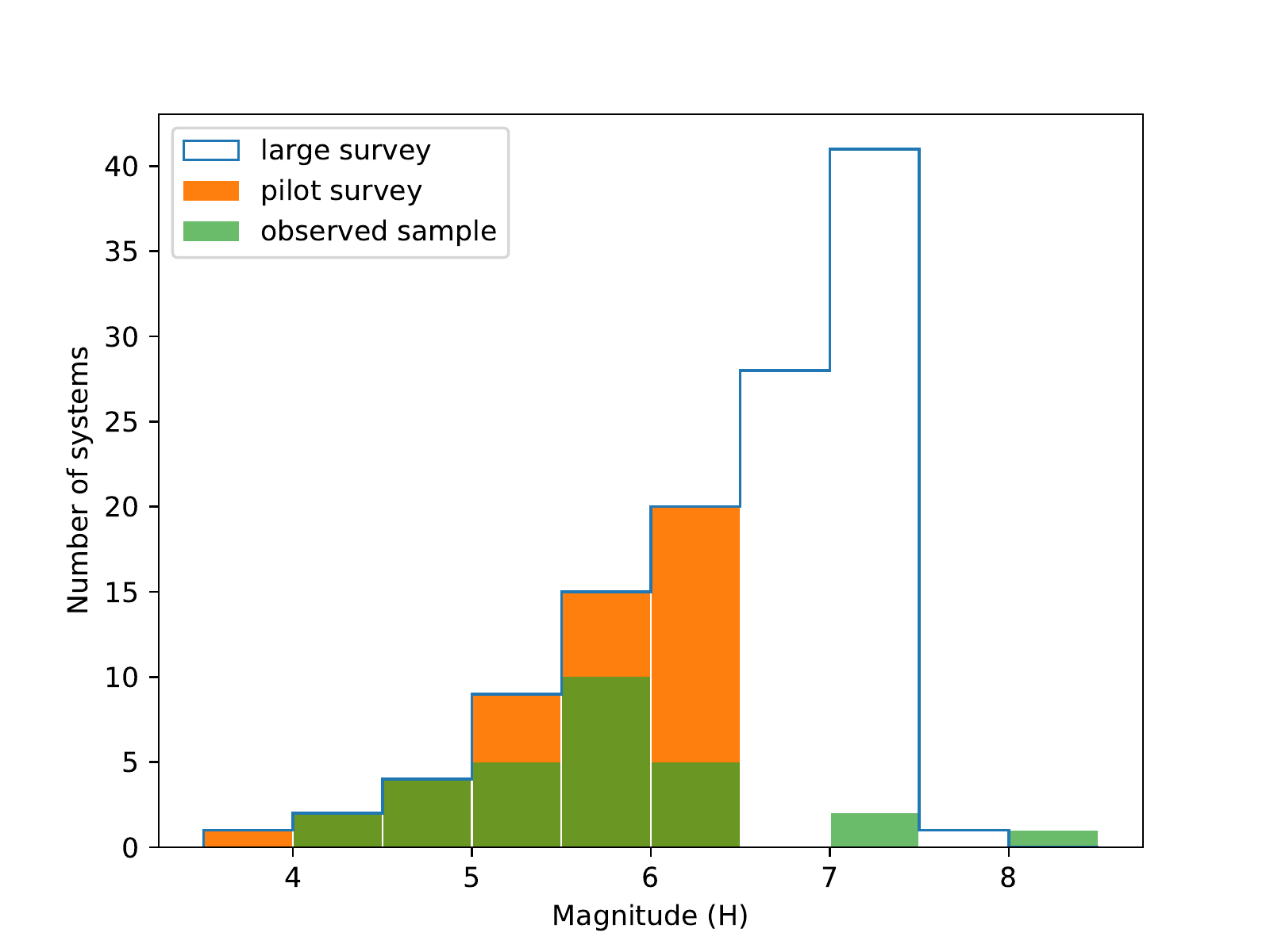}
\caption{Histogram of the H-band magnitude of our observed sample in green, the original sample planned for this pilot survey in orange, and the planned sample for the large survey in blue.} \label{fig:Hmag_hist}
\end{figure}
Figure~\ref{fig:Hmag_hist} shows the histogram of O-type star systems for which we obtained good-quality data as a function of their magnitude in the H-band. As the official magnitude limit in typical conditions is $H = 6.5$~mag, the bulk of observed objects is around this limit. But as we advanced in time, the improved knowledge of the MIRC-X instrument acquired during commissioning allowed us to push for fainter targets, beyond the initial magnitude limit of the instrument. Ultimately, observing an O-type star with a magnitude up to $H = 8.1$~mag was possible in excellent seeing conditions, as well as a couple of systems around $H = 7.5$~mag in normal seeing conditions. These latter observations are important to demonstrate that one can observe a large number of O-type stars with MIRC-X. Indeed, as shown in Fig.~\ref{fig:OvsHmag_cat}, we can observe up to 120 systems if a magnitude limit of $H = 7.5$~mag can be reached.

Usually, to calibrate a science target with the OLBI technique, we use "CAL1-SCI-CAL2" sequences, meaning that we observe a first calibrator (CAL1), then the science target (SCI), and finally a second calibrator (CAL2). To choose CAL1 and CAL2, we take unresolved targets within a reasonable distance and a similar apparent brightness to the science target, as explained above. It appears that two consecutive science targets observed in this program are close enough in the sky and in magnitude that CAL2 of one science target can be used as CAL1 of the next science target. Therefore, we chose to follow an observing sequence such as "CAL1-SCI1-CAL2-SCI2...," using each best-suited calibrator for each science target. This allows us to reduce the error due to calibration. When the same calibrator was best suited for two science targets observed one after the other (ex: SCI1 and SCI2), we preferred to use another calibrator for the second science target instead of using the same calibrator for both science targets (CAL1-SCI1-CAL1-SCI2). This reduces the risk that the results might be affected by a bad calibrator. A list of the calibrators observed but discarded because suspected to be actually multiple systems can be found in Appendix~\ref{Ann:cal}.

\subsection{Data reduction}\label{subsec:datared}

\begin{figure*}[!ht]
\centering
\includegraphics[scale=0.32]{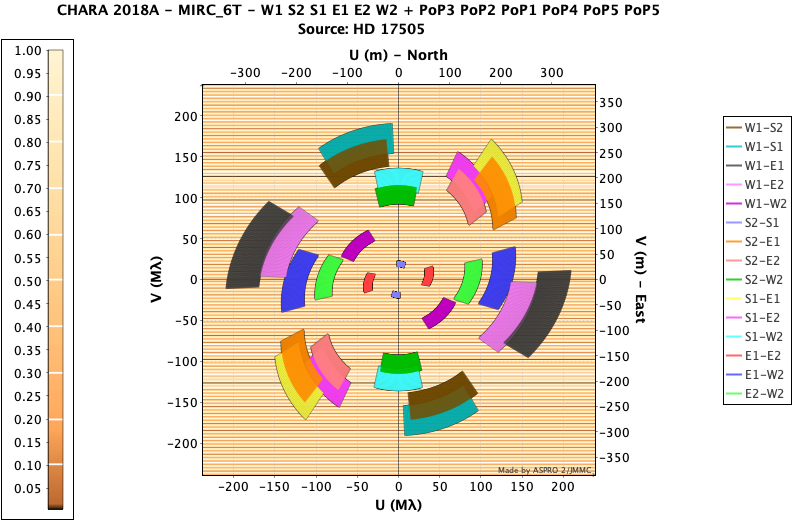}
\includegraphics[scale=0.32]{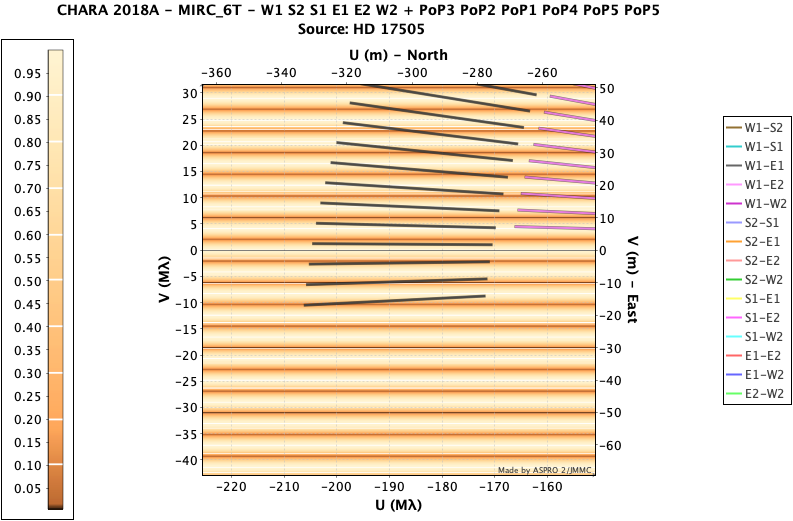}
\caption{Example of change in the uv-plane modeled with the ASPRO2 software. Left: global view of the uv-plane. Different colors are for different pairs of telescopes (baselines). Right: zoom on the change in the uv-plane for the fastest moving baseline. Each straight black line indicates the uv-coordinate for this baseline for an instant snapshot, and each snapshot is separated by 5 minutes. The regularly spaced, orange lines display the V2 values modeled for a binary with a separation of 50~mas.} \label{fig:aspro}
\end{figure*}

We used the MIRC-X data reduction pipeline\footnote{\href{https://gitlab.chara.gsu.edu/lebouquj/mircx_pipeline}{https://gitlab.chara.gsu.edu/lebouquj/mircx\_pipeline}}~\citep{anugu2020}. We set the maximum integration time (max-integration-time keyword in the pipeline) for one calibrated file to 220 seconds. The max-integration-time set the maximum time one reduced file will cover, with one file giving one constituent of the calibrated parameters. This means that one file will give the average calibrated parameters over the given maximum integration time. This time allows us to bin our data recording sequences of about 10 minutes into three files with a similar signal-to-noise ratio (S/N). This is needed because the change of the uv-coordinates due to the rotation of the Earth affects the observed squared visibility (V2). V2 can indeed change significantly in a timescale of 300 seconds for binaries at the limit of the OWA (see Fig.~\ref{fig:aspro} for an example).

To get rid of some outlier points due to bad S/N data, we applied a threshold on the S/N of four instead of the default value of 3. This limit allows us to select only data with high enough quality, while at the same time preserving as much data as possible.
To increase the quality of the reduced data, one would like to coherently sum the data over as long a time as possible. However, the atmospheric conditions limit the amount of time one can coherently integrate, as the phase induced by the atmosphere blurs the fringes, therefore reducing the quality of the data. The optimum time interval over which we can sum data for V2 and closure phase (CP) can however be different. To find this optimum value, we reduced the data for different values of the coherent integration time (ncoherent keyword in the reduction pipeline), which is the number of frames we add together to increase photon S/N, a frame being one recorded image of the detector. The pipeline produces a measurement of the calibrated parameters for each of these coherently added frames, and it is these coherently added frames that are incoherently averaged in the final calibrated file. Then, for each baseline and each target, we plotted the mean value over the files of the V2 S/N and the CP error as a function of the coherent integration time. We finally merged the V2 of the files with the higher S/N on V2 and the CP of the files with the lower error on CP, for most of the baselines or triplets. For the V2 we chose the same coherent integration time throughout the night. For the CP we took the best coherent integration time for each target. The number of coherently added frames for every target is specified in Table~\ref{tab:cohtime}, each frame integrating the flux for about 3~ms.

The data reduction pipeline provides two options to unbias the CP. We chose the option that computes an estimation of the bias as it does not require tuning any extra parameters (in opposition to the second method). This unbiasing method is explained in Appendix B of \cite{anugu2020}.

\section{Data analysis}\label{sec:dataanal}

To analyze the reduced data, and look for companions in the interferometric signal, we used the CANDID\footnote{\href{https://github.com/amerand/CANDID}{https://github.com/amerand/CANDID}} software~\citep{CANDID}. CANDID is a tool developed to look for binarity signals in interferometric data as well as to determine the position and flux ratio of the companion(s). For nondetections, CANDID also provides upper limits on the contrast of potential companions. More information about the algorithms and methods can be found in \citet{CANDID}.
To streamline the analysis of the data given the relatively large size of our sample, we established an automated procedure to analyze all the data consistently and uniformly. 

This procedure is summarized here.
First, for each observation of a star, we input all the reduced data to CANDID. 
We then perform a first search for a companion, fixing the maximum separation to 50~mas, which is approximately our OWA (see Eq.\ref{eq:OWA}). We fixed the step of the search grid to 1.0~mas. This step size is fine enough to find the global minimum in the range of separation we are looking for, while it is large enough to preserve reasonable computation time (approximately an hour per observation).
If a companion is detected (n$\sigma$ >5, with n$\sigma$ being the significance of the binary model compared to a uniform disk model fitting the data), we use the bootstrap function of CANDID around the position of the found companion. This gives us errors on the position and flux ratio of the companion that are more realistic than those computed by the initial grid search method.
Then, if a first companion is detected, we analytically remove the signal of this first companion from the interferometric data, and we perform a search for a second companion, using the same parameters as for the first search. We note that the detection of a second companion is made with an indirect method assuming the signal of the first companion is perfectly analytically removed, with no residual. Therefore, these detections should be taken with caution and would need further observations to confirm them.
Finally, we determine the upper limiting $\Delta$magnitude of a companion detectable in the data. If no companion is detected, we compute this limit on the reduced data directly. It gives us the limit in $\Delta$magnitude for which we would detect a companion. If a first companion is detected, we perform this computation on the data with the signal of the first companion analytically removed. It then gives us the limit in $\Delta$magnitude for which we would detect a second companion. We note that we can only remove the signal of one companion, so, the results of this detection limit might be biased if the signal of a second companion is present in the data.

Given the distance of these systems, even optical interferometry with 330~m baselines is not resolving the diameter of the stars in those systems. Hence, we consider them as a point source in CANDID, reducing the number of free parameters in the search for companions.

\section{Results}\label{sec:Resdet}

A summary of the results is shown in Table~\ref{tab:detect} for the first companion search and Table~\ref{tab:detect2} for the search of a second companion, with parameters that characterize the systems.

\subsection{New detections}\label{subsec:1stdet}
We here summarize the newly detected companions.

\textbf{Cyg OB2-5 A / BD+40 4220A / Schulte 5:} This system includes a short 6.6~d period binary, along with other components on wider orbits \citep{1999ApJ...517..416R,2019A&A...627A...2R}. Two distant companions were already known at separations of 0.93\arcsec~and 5.55\arcsec~\citep{2010A&A...518A...1M,2014AJ....147...40C,2020AJ....160..115C}, which is largely outside the OWA of our observations. Cyg OB2-5 A is a known PACWB, with the nonthermal radio emission mainly associated with a wide orbit with a period of about 6.7 years \citep{2010ApJ...709..632K}. We detect a companion in both observations of June 2018 and June 2019, at a mean separation of 12.18$\pm 0.35$~mas and a mean $\Delta H = 1.63 \pm 0.3$~mag, and a second companion in the observation of June 2019 at a separation of 6.51$\pm 0.26$~mas and a $\Delta H = 4.13 \pm0.01$~mag. The computed detection limit for the search of a second detection for the observation of June 2018 is 4 magnitudes fainter than the primary component, while the detected second companion in June 2019 is 4.13 magnitudes fainter. This could explain why we did not detect it in the June 2018 data, but further observations are necessary to confirm this second companion. This is the first direct detection for both companions. The wind-wind collision between the combined wind of the inner system and the one of the newly detected companions is likely the cause of the synchrotron radio emission. Long-term monitoring should allow us to determine which is associated with the 7-year period.

\textbf{Cyg OB2-10 / 2MASS J20334610+4133010 / Schulte 10:} Three companions have already been observed previously at separations of 0.21\arcsec, 0.74\arcsec~and 4.16\arcsec~with $\Delta K = 2.80\pm0.78$, $5.24\pm0.05$, and $6.03\pm0.07$~mag \citep{2020AJ....160..115C}. In the observation of June 2018, we detected a companion at a separation of 7.35$\pm 0.20$~mas and $\Delta H = 2.45 \pm0.03$~mag, but we do not detect it again in the observation of June 2019. More observations could be useful to confirm this new companion detection.


\textbf{HD~17505:} This object is known as a hierarchical triple system~\citep{2006ApJ...639.1069H, 2014ApJS..211...10S}. The system includes a close binary with an orbital period of 8.571 days, separated from the primary component by 2.161" \citep{2019A&A...626A..20M}. We detect an additional companion at a separation of 15.43$\pm 0.02$~mas from the primary component and $\Delta H = 0.35 \pm 0.04$~mag in two different observations separated by 1 day. This companion is detected for the first time, making this system a quadruple one.

\textbf{HD~19820:} This system is a known spectroscopic binary (SB) with a period of 3.36 days~\citep{1994A&A...282..455H}. We detect a companion at a mean separation of 13.87$\pm 0.03$~mas and $\Delta H = 2.57 \pm 0.01$~mag in two different observations separated by 1 day, and another companion at a separation of 6.96$\pm 0.11$~mas and $\Delta H = 4.16 \pm 0.01$~mag in only one of these observations. The detection of this second companion is made with an indirect method assuming the signal of the first companion is perfectly analytically removed, with no residual. Therefore, this detection should be taken with caution and will necessitate further observations to be confirmed.

\textbf{HD~24431:} This system has a companion situated at 0.72" \citep{mason, 2008AJ....136..554T}. We detect a new companion with a separation of 9.74$\pm 0.02$~mas and $\Delta H = 1.37 \pm 0.01$~mag.

\textbf{HD~28446:} This system is supposed to be an SB2 (double line spectroscopic binary) with a period of 1.31 days in~\cite{1994CoSka..24...65M} but has not been confirmed since. A third component is known at a separation of 10" \citep{2008MNRAS.389..869E}. We detect a fourth companion in both our observations of February 2018 and June 2018 with a mean separation of 26.16$\pm 0.13$~mas and a mean $\Delta H = 1.35 \pm 0.08$~mag.

\textbf{HD~34078:} This star is a known runaway \citep{2001A&A...365...49H}. Candidate companions have been detected by direct imaging at a separation of 8.4\arcsec~\citep{mason} and 0.35\arcsec~\citep{2008AJ....136..554T}, but both those detections are suspected to be field stars observed in the line of sight of this star. We detected a new companion in two different observations, at a separation of 6.85$\pm 0.07$~mas and $\Delta H = 2.76 \pm 0.02$~mag in December 2017 and a separation of 1.74$\pm 0.20$~mas and $\Delta H = 3.29 \pm 0.03$~mag in September 2018.

\textbf{HD~36861:} This star ($\lambda$ Ori A) is known to be variable~\citep{1996ApJS..103..475F}, and part of a wide binary (HD~36861J). The components A and B of the system $\lambda$ Ori may not be physically bound~\citep{1985A&AS...60..183L, mason}. $\lambda$ Ori A had no detected companion yet~\citep{2009AJ....137.3358M}, but we detect a companion in our observation of February 2018, with a separation of 10.13$\pm 0.05$~mas and a $\Delta H = 3.30 \pm 0.02$~mag. This companion has not been detected again in our observation of June 2018, despite the computed limiting magnitude of detection being $\Delta H = 5.03$~mag. The reason for this nondetection on the second observation is unknown. A new observation of this system could confirm this new companion.

\textbf{HD~47129:} This is Plaskett's star. It is a known SB2 with a period of 14.4 days \citep{2008A&A...489..713L}. There are also two known visual companions at 0.78" and 1.12" \citep{2008AJ....136..554T}. The SMASH+ survey resolved a faint companion at 36 mas with $\Delta H \sim 4.0$~mag, with the NACO/SAM instrument, but it was too faint to be confirmed by PIONIER and the uncertainties on the separation found by NACO/SAM were large. We detect a companion at a separation of 32.29$\pm 0.06$ mas and $\Delta H = 4.6\pm 0.01$~mag. This detected companion is a confirmation of the candidate by NACO/SAM, providing compelling evidence for the existence of this companion. Interestingly enough, this long-period companion is compliant with the likely membership of this system to the category of PACWBs, based on radio results published by \citet{2017MNRAS.465.2160K}.

\textbf{HD~207198:} This system has a known companion at a separation of 17.64\arcsec~\citep{2004AJ....128.3012M}. We detect for this system a companion at a separation of 41.07$\pm 0.04$~mas and $\Delta H = 4.68 \pm 0.01$~mag. This companion is detected for the first time. We note that this detection is close to the OWA.

\textbf{HD~229196:} We detect for the first time in this system a companion, situated at a separation of 5.88$\pm 0.02$~mas and a mean $\Delta H = 2.80 \pm 0.04$~mag in two different observations separated by 1 day.

\subsection{Already detected}\label{subsec:alrdet}
We here summarize the redetection of companions that were already known from other techniques or previous optical interferometric observations.

\textbf{Cyg OB2-9 / HIP 101419:} This PACWB is a known very excentric SB2~\citep{2013A&A...550A..90B,2019A&A...626A..20M,2020AJ....160..115C} with a period estimated at 2.4 years and an excentricity of 0.713~\citep{2010ApJ...719..634N,2012A&A...546A..37N} with another companion at a separation of 21\arcsec~\citep{2010A&A...518A...1M}. We detect a companion at a mean separation of 0.77$\pm 0.01$~mas and a mean $\Delta H = 0.42 \pm 0.04$~mag in two observations separated by two days. With the orbital parameters in \citet{2013A&A...550A..90B} we derive a minimum projected separation of $\sim2.04$~AU with the known SB2 companion. Taking the distance of the system and the angular separation we detect our companion, we compute a projected separation of $\sim 1.38$~AU. Taking the uncertainties into account, our detection is probably the SB2 component observed close to its periastron. Further observations could confirm it.

\textbf{HD~47839 / 15 Mon:} This system has a known companion at $\simeq$ 0.1\arcsec~with a difference of magnitude of 1.6 in the visible~\citep{2021ApJS..257...69H} and a third component at a wider (3~\arcsec) separation \citep{mason, Sana2014}. We detect a companion with a separation of at least 49.19$\pm 0.32$~mas and $\Delta H = 1.81 \pm 0.01$~mag. This detection is at the limit of the OWA, which means that it is probable that the companion detected is further out, as discussed in Sect.~\ref{disc:OWA}. This companion is probably the known companion around 100~mas, as the differences in magnitude are compatible and as the separation close to the OWA cannot rule out this known companion.

\textbf{HD~167971:} This system is a known hierarchical triple system \citep{1987A&A...185..121L,2012MNRAS.423.2711D, 2017A&A...601A..34L, 2019A&A...624A..55S}, that also turns out to be the brightest synchrotron-emitting O-type PACWB in the catalog \citep{2013A&A...558A..28D}. The central binary has an orbital period of 3.3 days and the third component is orbiting the inner binary on a timescale of 21.4 years. We detect a companion at 19.89$\pm 0.01$~mas and $\Delta H = 0.61 \pm 0.01$~mag. The separation of the detected companion is compatible with the one of the outer component of the system measured in \citet{2012MNRAS.423.2711D} and \citet{2017A&A...601A..34L}. 

\textbf{HD~193322:} This is a complex multiple system. The A component consists of a single star Aa orbiting around a 312-day binary Ab in 35~years \citep{2011AJ....142...21T}. Three other components are also known, with a separation of 2.6\arcsec~for the closest \citep{2008AJ....136..554T}. We detect a companion at a separation of at least 47.33~mas and maximum $\Delta H = 0.06 ^{+0.06}_{-0.05}$~mag in our three observations of this system in June 2018, in September 2018, and in June 2019. These detections are at the limit of the OWA, which means that it is possible that the companion is somewhat further out, as discussed in Sect.~\ref{disc:OWA}. The detected companion is most likely the already known pair Aa and Ab due to its separation being compatible with the known pair reported in \citet{2011AJ....142...21T}.

\textbf{HD~202214:} One close companion is already known from spectroscopy, with an orbital period of 81.30 days \citep{2002IAUDS.147....2M}. The system also has two wider companions, one at a separation of 0.071\arcsec~\citep{2002IAUDS.147....2M} and a $\Delta V = 0.6$~\citep{2009AJ....137.3358M}~mag and one at 1.0\arcsec~with a $\Delta V = 0.3$~\citep{2009AJ....137.3358M}~mag. We detect a companion at a separation of 47.27$\pm 0.05$~mas and $\Delta H = 0.03 ^{+0.03}_{-0.04}$~mag. This detection is at the limit of the OWA, which means that it is probable that the detected companion is actually further out, as discussed in Sect.~\ref{disc:OWA}. We, therefore, cannot rule out that the detected companion is the known one at 71~mas. The difference between our magnitude difference and the one from~\citet{2009AJ....137.3358M} could then be explained by the separation potentially being larger than our OWA, introducing bias in our detection, and by the fact that our observing setup is not optimized for companions outside of the OWA. The derived contrast might therefore be systematically biased.

\textbf{HD~206267:} This system is a high order multiple system \citep{2019A&A...626A..20M, 2020A&A...636A..28M}. The central component (AaAb) is composed of an SB2 (Aa) with a period of 3.71 days \citep{2018A&A...614A..60R} and another companion (Ab) separated from Aa by 0.1\arcsec~and a difference in magnitude of 1.63$\pm$0.3 at a wavelength of $\lambda = 0.91~\mu$m (SDSS z filter\footnote{\href{https://skyserver.sdss.org/dr1/en/proj/advanced/color/sdssfilters.asp}{https://skyserver.sdss.org/dr1/en/proj/advanced/color/sdssfilters.asp}}) \citep{2020A&A...639C...1M}. A third component (B) is situated at a separation of 1.7\arcsec~with a difference of 5.72$\pm$0.13 in magnitude at $\lambda = 0.91~\mu$m. Two other companions (C and D) are situated within 25\arcsec. We detect a companion at a separation of at least 49.52$\pm 0.22$~mas with a difference of 1.64$\pm 0.02$~mag in the H-band. This detection is at the limit of the OWA, which means that it is probable that this detected companion is further out, as discussed in Sect.~\ref{disc:OWA}. We, therefore, cannot rule out that this detection can be the known companion at 0.1\arcsec; the magnitude differences are compatible one with each other.

\subsection{No detection}\label{subsec:nodet}

We do not detect interferometric companions around HD~30614~\citep[SB1 system with a period of 3.68 days; ][]{1986SvAL...12..125Z}, HD~47432 \citep[one known companion at 0.78\arcsec; ][]{2008AJ....136..554T}, HD~188001~\citep[single runaway star;][]{2021A&A...655A...4T}, HD~195592  \citep[spectroscopic binary with a period of a few days;][]{2010NewA...15...76D}, HD~201345~\citep[one known wide companion at 7.38\arcsec; ][]{2008AJ....136..554T}, HD~206183~\citep[two known wide companions at separations of 6.6\arcsec~and 11.6\arcsec; ][]{mason}, HD~209975~\citep[four known wide companions at 3.79\arcsec, 4.14\arcsec, 19.8\arcsec, and 60.4\arcsec; ][]{mason, 2008AJ....136..554T}, HD~217086 \citep[2 known companions at 2.8\arcsec~and 3.1\arcsec; ][] {mason, 2008AJ....136..554T}, and HD~228779.

\subsection{Candidates}\label{subsec:candidates}

Some of our observations resulted in candidates with only a marginally significant detection criterion ($3 < n\sigma < 5$). Those systems are HD~45314~\citep[Oe star][]{2015A&A...575A..99R}, HD~210809, and HD~210839. More observations will be needed to validate or reject the presence of a companion. The position of those candidates can be found in Table~\ref{tab:detect}.

\begin{table*}[ht!]
\caption{Summary of the results for the first companion search with CANDID.}
\label{tab:detect}
\centering
\begin{tabular}{cccccccccc}\hline\hline
Target's name & DATE-OBS & n$\sigma$  & sep & P.A. & emax &  emin & P.A. emax & $\Delta H$ & det. lim. \\
&MJD&&[mas]&[deg]&[mas]&[mas]&[deg]&[mag]&$\Delta$mag(H)\\\hline

Cyg OB2-5 A & 58279.353 & 6.46 & 13.35 & 97.45 & 0.35 & 0.11 & $-3.03$ & 1.91 $^{+ 0.02 }_{- 0.02 }$ & -- \\ 
... & 58657.396 & 8.03 & 11.00 & 107.12 & 0.00 & 0.00 & 26.56 & 1.38 $^{+ 0.00 }_{- 0.00 }$ & -- \\ 
Cyg OB2-9 & 58386.154 & 8.03 & 0.76 & 148.90 & 0.01 & 0.00 & $-13.26$ & 0.38 $^{+ 0.02 }_{- 0.02 }$ & -- \\ 
... & 58388.197 & 8.03 & 0.78 & 152.99 & 0.01 & 0.01 & 81.50 & 0.45 $^{+ 0.01 }_{- 0.01 }$ & -- \\ 
Cyg OB2-10 & 58281.491 & 5.75 & 7.35 & $-21.47$ & 0.20 & 0.03 & 38.96 & 2.45 $^{+ 0.03 }_{- 0.03 }$ & -- \\ 
... & 58657.446 & 2.96 & -- & -- & -- & -- & -- & -- & 3.24 \\ 
HD~17505 & 58386.393 & 8.03 & 15.43 & $-141.24$ & 0.02 & 0.01 & 40.24 & 0.35 $^{+ 0.04 }_{- 0.03 }$ & -- \\ 
... & 58387.358 & 8.03 & 15.43 & $-141.22$ & 0.01 & 0.00 & 52.60 & 0.34 $^{+ 0.02 }_{- 0.03 }$ & -- \\ 
HD~19820 & 58386.331 & 8.03 & 13.89 & 82.35 & 0.03 & 0.01 & 1.55 & 2.57 $^{+ 0.01 }_{- 0.01 }$ & -- \\ 
... & 58387.424 & 8.03 & 13.85 & 82.56 & 0.02 & 0.02 & 36.70 & 2.56 $^{+ 0.01 }_{- 0.01 }$ & -- \\ 
HD~24431 & 58385.388 & 49.96 & 9.74 & 146.19 & 0.02 & 0.01 & $-82.80$ & 1.37 $^{+ 0.01 }_{- 0.00 }$ & -- \\ 
HD~28446 & 58157.339 & 47.82 & 25.06 & $-45.64$ & 0.13 & 0.02 & $-86.04$ & 1.42 $^{+ 0.02 }_{- 0.01 }$ & -- \\ 
... & 58385.363 & 49.96 & 27.25 & $-56.15$ & 0.01 & 0.01 & $-52.32$ & 1.28 $^{+ 0.01 }_{- 0.01 }$ & -- \\ 
HD~30614 & 58386.439 & 1.34 & -- & -- & -- & -- & -- & -- & 4.32 \\ 
HD~34078 & 58100.407 & 5.62 & 6.85 & 172.83 & 0.07 & 0.06 & 77.06 & 2.76 $^{+ 0.02 }_{- 0.02 }$ & -- \\ 
... & 58384.400 & 5.29 & 1.74 & 171.17 & 0.20 & 0.04 & $-61.50$ & 3.29 $^{+ 0.03 }_{- 0.03 }$ & -- \\ 
HD~36861 & 58156.175 & 6.51 & 10.13 & $-11.06$ & 0.05 & 0.04 & 59.92 & 3.3 $^{+ 0.02 }_{- 0.02 }$ & -- \\ 
... & 58387.479 & 0.96 & -- & -- & -- & -- & -- & -- & 5.03 \\ 
HD~45314 & 58386.497 & 3.25 & 27.80 & 167.42 & 0.09 & 0.04 & $-21.81$ & 4.24 $^{+ 0.00 }_{- 0.01 }$ & -- \\ 
HD~47129 & 58385.529 & 5.01 & 32.39 & 37.50 & 0.06 & 0.03 & $-73.35$ & 4.6 $^{+ 0.01 }_{- 0.01 }$ & -- \\ 
HD~47432 & 58387.523 & 0.66 & -- & -- & -- & -- & -- & -- & 4.91 \\ 
... & 58388.512 & 0.00 & -- & -- & -- & -- & -- & -- & 4.99 \\ 
HD~47839 & 58386.535 & 17.76 & 49.19$^\ast$ & $-72.92$ & 0.32 & 0.17 & 86.62 & 1.81 $^{+ 0.01 }_{- 0.01 }$ & -- \\ 
HD~167971 & 58658.289 & 49.96 & 19.89 & $-98.40$ & 0.01 & 0.00 & $-20.18$ & 0.61 $^{+ 0.01 }_{- 0.01 }$ & -- \\ 
HD~188001 & 58387.135 & 2.23 & -- & -- & -- & -- & -- & -- & 4.25 \\ 
HD~193322 & 58281.321 & 49.96 & 47.33$^\ast$ & 15.78 & 0.24 & 0.13 & 10.57 & 0.06 $^{+ 0.06 }_{- 0.05 }$ & -- \\ 
... & 58388.146 & 49.96 & 48.70$^\ast$ & 148.78 & 0.27 & 0.08 & $-17.21$ & 0.01 $^{+ 0.1 }_{- 0.05 }$ & -- \\ 
... & 58658.367 & 49.96 & 49.18$^\ast$ & 148.12 & 0.11 & 0.05 & $-25.95$ & $-0.01 ^{+ 0.00 }_{- 0.00 }$ & -- \\ 
HD~195592 & 58280.337 & 1.89 & -- & -- & -- & -- & -- & -- & 3.63 \\ 
HD~201345 & 58661.420 & 2.82 & -- & -- & -- & -- & -- & -- & 4.38 \\ 
HD~202214 & 58658.412 & 49.96 & 47.27$^\ast$ & $-129.68$ & 0.05 & 0.04 & 71.07 & 0.03 $^{+ 0.03 }_{- 0.04 }$ & -- \\ 
HD~206183 & 58661.470 & 0.67 & -- & -- & -- & -- & -- & -- & 5.83 \\ 
HD~206267 & 58660.442 & 8.03 & 49.52$^\ast$ & $-145.71$ & 0.22 & 0.06 & 0.38 & 1.64 $^{+ 0.02 }_{- 0.01 }$ & -- \\ 
HD~207198 & 58658.449 & 7.10 & 41.07$^\ast$ & $-32.93$ & 0.04 & 0.02 & $-34.19$ & 4.68 $^{+ 0.00 }_{- 0.01 }$ & -- \\ 
HD~209975 & 58278.401 & 0.76 & -- & -- & -- & -- & -- & -- & 5.16 \\ 
HD~210809 & 58660.490 & 3.14 & 4.45 & 70.72 & 0.09 & 0.04 & $-64.87$ & 4.02 $^{+ 0.01 }_{- 0.02 }$ & -- \\ 
HD~210839 & 58657.486 & 4.93 & 6.72 & 79.36 & 0.08 & 0.02 & $-60.61$ & 5.92 $^{+ 0.00 }_{- 0.00 }$ & -- \\ 
HD~217086 & 58387.243 & 0.10 & -- & -- & -- & -- & -- & -- & 4.69 \\ 
HD~228779 & 58660.400 & 2.50 & -- & -- & -- & -- & -- & -- & 4.97 \\ 
HD~229196 & 58385.195 & 8.03 & 5.88 & 13.52 & 0.03 & 0.01 & 77.55 & 2.76 $^{+ 0.00 }_{- 0.00 }$ & -- \\ 
... & 58388.244 & 8.03 & 5.88 & 14.02 & 0.03 & 0.02 & 89.62 & 2.84 $^{+ 0.01 }_{- 0.01 }$ & -- \\ 

\hline
\end{tabular}
\tablefoot{The first column is the system name and the second column is the MJD date of the observations. The third column is the significance of the detection of a companion by CANDID. The fourth and fifth columns are respectively the separation and the position angle (P.A.) of the detected companion. The P.A. is defined as the angle between the north direction and the companion direction with respect to the central element counted positively toward the east. The separations noted with an $^{\ast}$ symbol are close to the OWA, and should only be considered as a lower limit. The sixth, seventh, and eighth columns give respectively the semi-major axis and the semi-minor axis, and the P.A. between the semi-major axis direction and the north direction of the error's ellipse on the position of the detected companion. The ninth column is the magnitude difference between the main component and the detected companion in the H-band. The tenth column is the $\Delta$magnitude limit for which we should be able to detect a companion computed by CANDID. We note that for the systems with no companion detected, we only give the result of the detection limiting magnitude, and we do not give them for the systems with candidates or detected companions, as we give them in Table~\ref{tab:detect2}.}
\end{table*}

\begin{table*}[ht!]
\caption{Same as Table~\ref{tab:detect} but for the second detection.}
\label{tab:detect2}
\centering
\begin{tabular}{cccccccccc}\hline\hline
Target's name & DATE-OBS & n$\sigma$  & sep & P.A. & emax &  emin & P.A. emax & $\Delta$H & det. lim. \\
&MJD&&[mas]&[deg]&[mas]&[mas]&[deg]&[mag]&$\Delta$mag(H)\\\hline

Cyg OB2-5 A & 58279.353 & 0.74 & -- & -- & -- & -- & -- & -- & 4.06 \\ 
... & 58657.396 & 8.03 & 6.51 & $-85.06$ & 0.26 & 0.04 & 85.05 & 4.15 $^{+ 0.01 }_{- 0.01 }$ & -- \\ 
Cyg OB2-9 & 58386.154 & 0.76 & -- & -- & -- & -- & -- & -- & 4.98 \\ 
... & 58388.197 & 0.67 & -- & -- & -- & -- & -- & -- & 5.88 \\ 
Cyg OB2-10 & 58281.491 & 1.62 & -- & -- & -- & -- & -- & -- & 3.51 \\ 
HD~17505 & 58386.393 & 3.72 & -- & -- & -- & -- & -- & -- & 4.19 \\ 
... & 58387.358 & 0.7 & -- & -- & -- & -- & -- & -- & 1.56 \\ 
HD~19820 & 58386.331 & 8.03 & 6.96 & 91.45 & 0.11 & 0.05 & 89.91 & 4.16 $^{+ 0.01 }_{- 0.01 }$ & -- \\ 
... & 58387.424 & 0.99 & -- & -- & -- & -- & -- & -- & 4.29 \\ 
HD~24431 & 58385.388 & 1.83 & -- & -- & -- & -- & -- & -- & 4.77 \\ 
HD~28446 & 58157.339 & 0.81 & -- & -- & -- & -- & -- & -- & 3.28 \\ 
... & 58385.363 & 1.81 & -- & -- & -- & -- & -- & -- & 4.72 \\ 
HD~34078 & 58100.407 & 4.74 & -- & -- & -- & -- & -- & -- & 3.33 \\ 
... & 58384.4 & 2.13 & -- & -- & -- & -- & -- & -- & 4.31 \\ 
HD~36861 & 58156.175 & 1.18 & -- & -- & -- & -- & -- & -- & 4.83 \\ 
HD~45314 & 58386.497 & 2.0 & -- & -- & -- & -- & -- & -- & 4.74 \\ 
HD~47129 & 58385.529 & 0.78 & -- & -- & -- & -- & -- & -- & 5.28 \\ 
HD~47839 & 58386.535 & 6.66 & -- & -- & -- & -- & -- & -- & 3.34 \\ 
HD~167971 & 58658.289 & 1.17 & -- & -- & -- & -- & -- & -- & 3.78 \\ 
HD~193322 & 58281.321 & 4.52 & -- & -- & -- & -- & -- & -- & 3.03 \\ 
... & 58388.146 & 2.33 & -- & -- & -- & -- & -- & -- & 3.16 \\ 
... & 58658.367 & 6.72 & -- & -- & -- & -- & -- & -- & 3.07 \\ 
HD~202214 & 58658.412 & 15.7 & -- & -- & -- & -- & -- & -- & 2.86 \\ 
HD~206267 & 58660.442 & 4.62 & -- & -- & -- & -- & -- & -- & 3.08 \\ 
HD~207198 & 58658.449 & 0.82 & -- & -- & -- & -- & -- & -- & 5.78 \\ 
HD~210809 & 58660.49 & 1.03 & -- & -- & -- & -- & -- & -- & 5.05 \\ 
HD~210839 & 58657.486 & 2.98 & -- & -- & -- & -- & -- & -- & 5.51 \\ 
HD~229196 & 58385.195 & 3.07 & -- & -- & -- & -- & -- & -- & 4.38 \\ 
... & 58388.244 & 1.24 & -- & -- & -- & -- & -- & -- & 4.42 \\ 

\hline
\end{tabular}
\tablefoot{The detection's limiting $\Delta$magnitude of systems with a second companion detected is not accurate as the data are polluted by the signal of the second companion.}
\end{table*}

\section{Statistical analysis}\label{sec:statanal}

Among the 29 systems observed with good-quality data, we confirm the detection of 19 companions for 17 multiple systems (see Sections~\ref{subsec:1stdet} and~\ref{subsec:alrdet}). Out of these 19 companions, 13 are detected for the first time. This gives us a multiplicity fraction $f_{\text{m}} = 17 / 29 = 0.59 \pm 0.09$, and a companion fraction of $f_{\text{c}} = 19 / 29 = 0.66 \pm 0.13$ in the range of separations to which we are sensitive. The uncertainty for the multiplicity fraction is obtained with a binomial uncertainty and the companion fraction uncertainty is obtained with a Poisson uncertainty.

We note that 38\% of our sample corresponds actually to multiple systems with at least three components and that in 28\% of the sample, the interferometric companion constitutes the detection of the outer orbit in a hierarchical triple system. This proportion of hierarchical triple systems should allow us to study the Kozai-Lidov effect~\citep{2016ARA&A..54..441N}, by studying their orbits.

\subsection{Comparison with SMASH+}\label{sebsec:compareSMASH}

The multiplicity fraction in this study is marginally higher than the one in SMASH+ ($0.41 \pm 0.05$). This difference could be explained by the fact that our preliminary magnitude limit ($H < 6.5$~mag) was brighter than the one of SMASH+ ($H < 7.5$~mag). This would push the observational bias to observe a larger fraction of multiple systems. 

By taking into account all known companions, summarized in Table~\ref{tab:logObs}, the total multiplicity fraction ($0.86 \pm 0.07$) and the total companion fraction ($2.10 \pm0.27$), are consistent with the one of SMASH+ (respectively $0.91 \pm 0.03$ and $2.1 \pm 0.2$).
We also find that in our sample, $38 \pm 9$\% of the systems contain spectroscopic binaries, which is marginally lower than in the SMASH+ sample ($49 \pm 5$\%).

\subsection{Estimated mass ratio distribution}\label{subsec:qdist}

We define the mass ratio $q$ as:
\begin{equation}
    q = M_{\text{comp}}/M_{\text{primary}} 
\end{equation}
where $M_{\text{comp}}$ is the mass of the detected companion and $M_{\text{primary}}$ is the mass of the primary component of the system.  This mass ratio can be estimated using the flux ratio between the two components. In this study, we use the same relation used in \cite{2017A&A...601A..34L} and derived from \cite{martins2005}, which gives a good approximation of the mass ratio for main-sequence stars:
\begin{equation}
    q = (f_{H})^{0.7}
\end{equation}
where $f_H$ is the flux ratio in the H-band, given by the CANDID analysis. We note that if the central component is an unresolved binary, this method of computing the mass ratio is using the combined flux of both components, hence the estimated mass ratio will be biased.

\begin{figure}[!ht]
\centering
\includegraphics[width=\columnwidth]{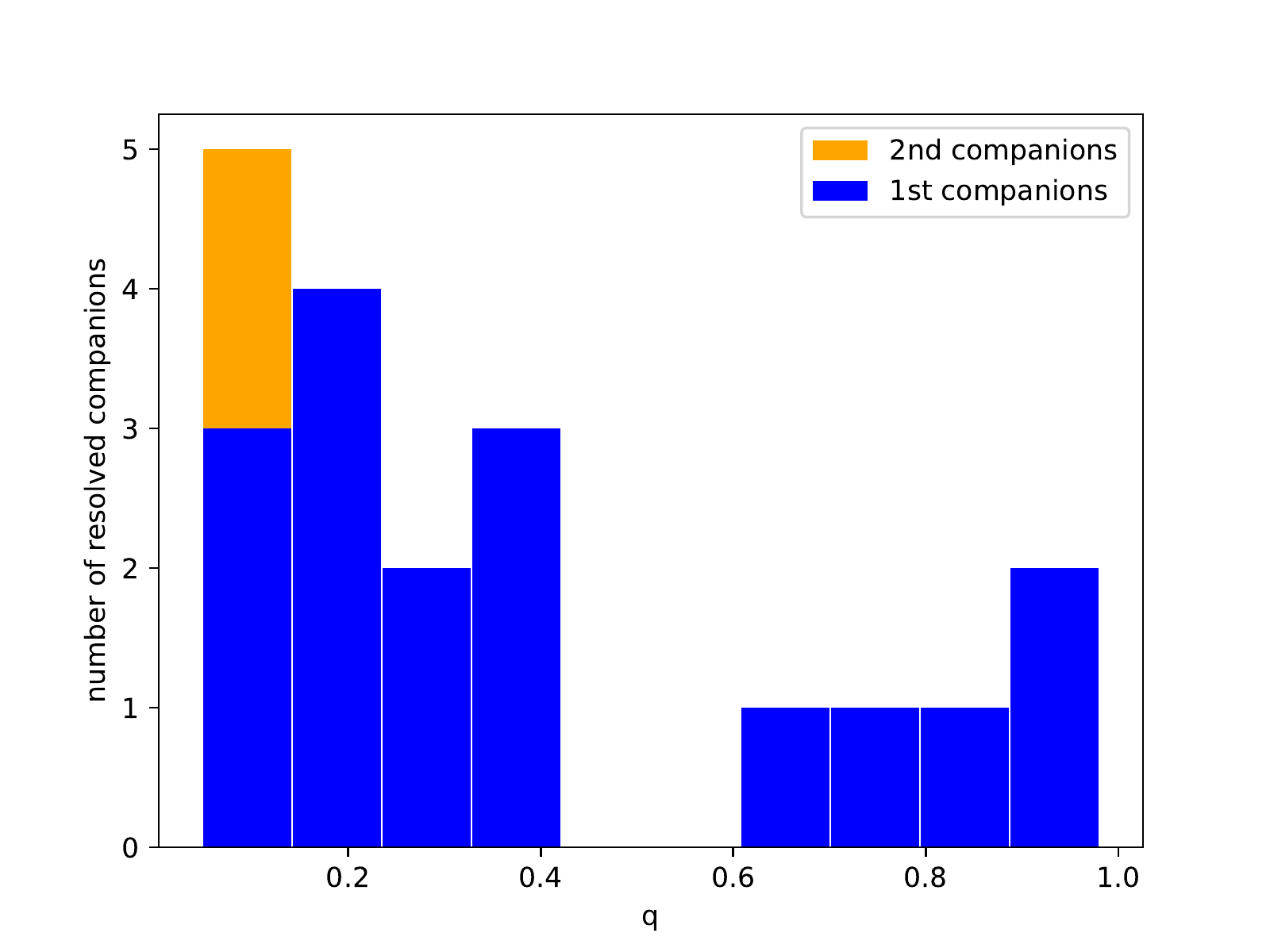}
\caption{Histogram of detected companions as a function of the estimated mass ratio $q$, when taking into account only the first detected companions (blue), and when adding the second companions detected (orange).} \label{fig:q_hist}
\end{figure}
Figure~\ref{fig:q_hist} displays the distribution of the estimated mass ratio of the detected companions. In this figure, we can see that the distribution seems to be bi-modal, with a lack of companions between $q = 0.4$ and $0.6$, and favoring a lower mass ratio. The bi-modality could be explained by the small statistic we are using here, with only 17 companions. However, we ran a Kuiper test to compare the estimated mass ratio distribution with a uniform distribution. The result of this test is a value of D $= 0.41$, and a probability of obtaining the value D from a uniform distribution of 2.7\%. From this low probability, we can conclude that the actual mass ratio distribution is most probably not uniform. 

The distribution favoring low mass ratios goes against the observational bias of the survey being magnitude limited. This bias should favor the inclusion of binaries with bright companions, therefore, with a flux ratio, hence a mass ratio, close to 1. So, this tendency seems to come from the intrinsic O-type stars' mass ratio distribution. The results of the survey of the massive stars in the Orion region \citep{ORION} show a similar trend, with a mass ratio distribution following a power law $\propto q^{\alpha}$ with $\alpha = 1.7$.

\subsection{Projected separation distribution}\label{subsec:sepAUdist}

In the absence of any estimate of the inclination of the orbit, the absolute physical separation cannot be determined. Rather, we determined the projected separation of the detected companions, using the distances published by \cite{2021yCat.1352....0B}. 
\begin{figure}[!ht]
\centering
\includegraphics[width=\columnwidth]{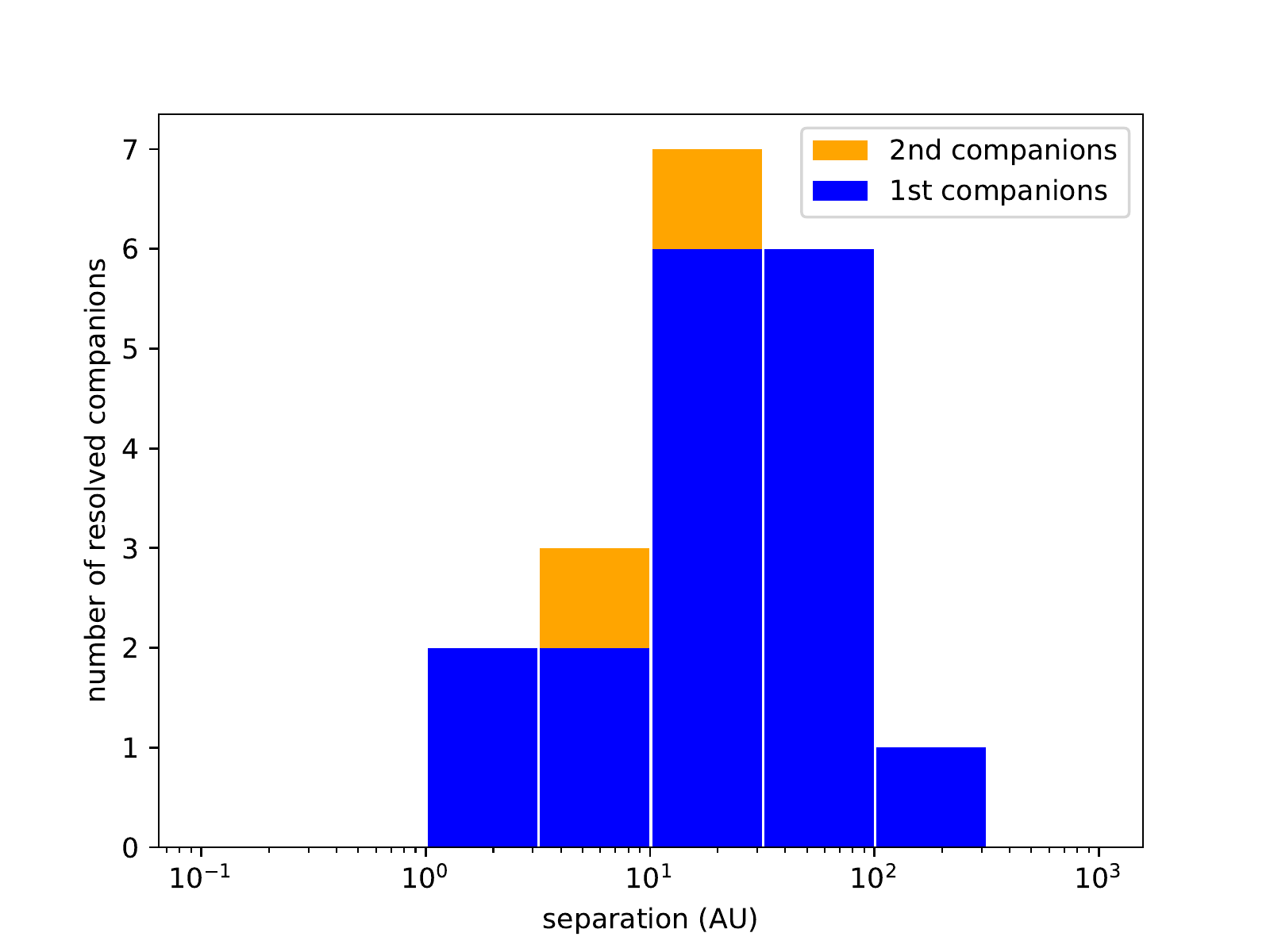}
\caption{Histogram of detected companions as a function of the projected separation in Astronomical Unit (AU). The distribution in blue takes into account only the first companions. The distribution in orange considers all companions.} \label{fig:sepAU_hist}
\end{figure}
Figure~\ref{fig:sepAU_hist} shows the distribution of the separation of the companions detected in this study. The distribution seems to favor the middle and high part of the probed separation, from 10 to 100 AU.
We note that the companions detected close to the OWA may actually be located further out, meaning that the real distribution could have a tail at larger separations. This figure only shows the companions that were (re)detected in this study.

\subsection{Estimated mass ratio as a function of the projected separation}\label{subsec:q_sepAU}

\begin{figure}[!ht]
\centering
\includegraphics[width=\columnwidth]{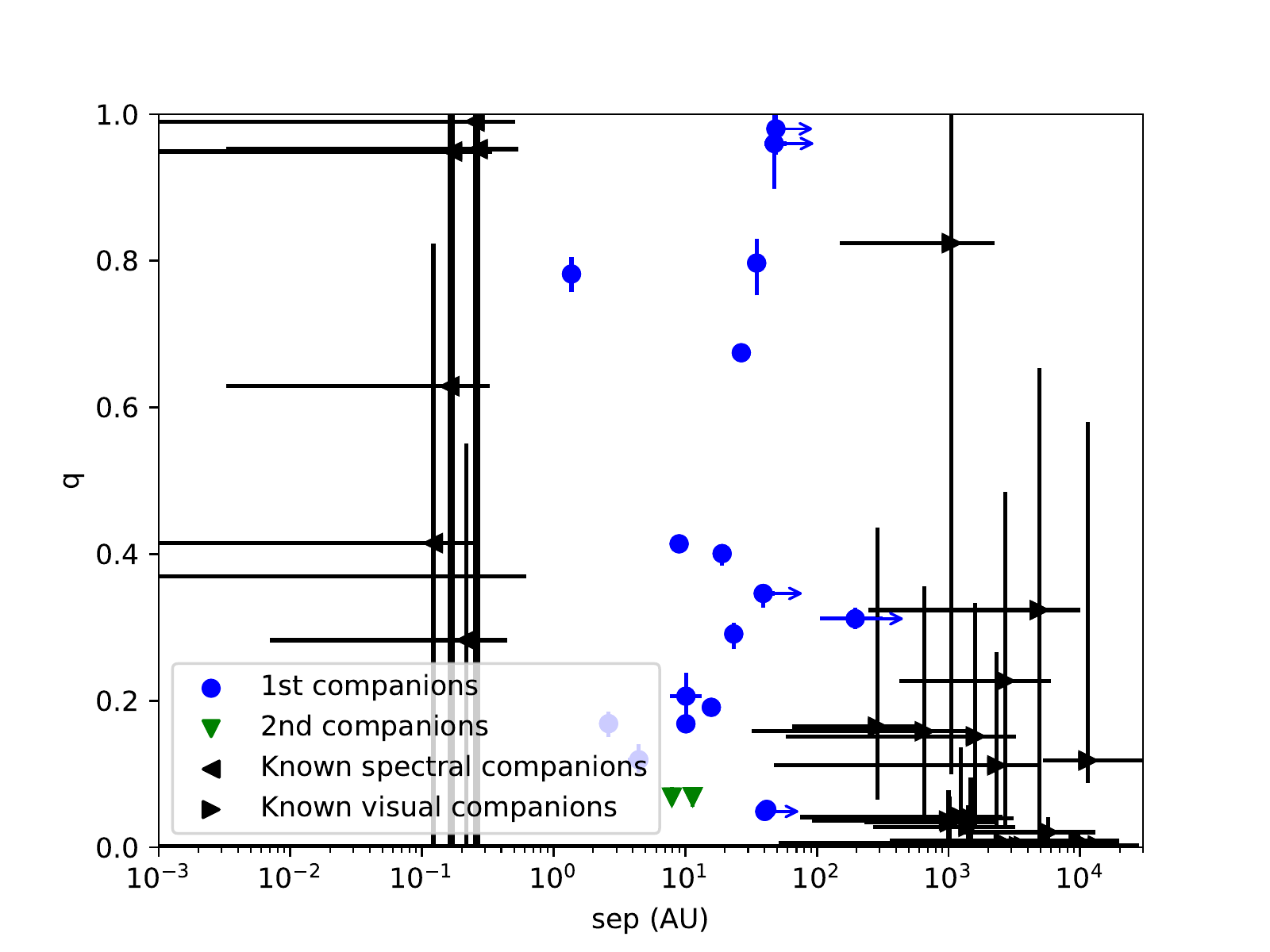}
\caption{Estimated mass ratio as a function of the projected separation in AU of the detected companions. We note that some error bars are hidden by the size of the markers. Error bars on separation ending with an arrow pointing right are to distinguish the companions detected close to the OWA. The black points are for known companions detected with other observational techniques.} \label{fig:q_sepAU}
\end{figure}
Figure~\ref{fig:q_sepAU} shows the estimated mass ratio $q$ as a function of the projected separation of the companions detected in this study. Error bars on separation ending with an arrow pointing right are to distinguish the companions detected close to the OWA, meaning that the physical separation on the plot is only a lower limit. The black points are for the companions detected with other techniques for which the mass ratio and physical separation are known or can be determined from the literature.

We notice the higher number of companions with an estimated mass ratio $q < 0.6$ as described in Sect.~\ref{subsec:qdist}. However, we cannot discern any correlations in this plot between the estimated mass ratio $q$ and the physical separation for the companions detected in this study as the companions seem relatively homogeneously distributed in the figure. 

Furthermore, taking into consideration the companions previously detected by other techniques, a correlation seems to appear, with $q$ being anticorrelated with the separation. This anticorrelation would be in favor of the competitive accretion formation process~\citep{ORION}. But one should note that this result could be due to the difference of biases in the detection of companions by the different techniques. A further study taking into account the different biases is necessary to bring a stronger conclusion.

We caution that the plot is affected by some factors that deserve a few comments. First of all, we certainly need a greater statistical sample if we want to be sure that there is or is not a correlation between these two parameters for massive star systems. For subsequent developments, we stress that we should ideally base our discussion on absolute physical separations, and not on projected ones, which requires knowledge about the inclination. This can be obtained through a suitable interferometric follow-up to derive the relative astrometric orbit of the systems. Finally, the best test for a potential (lack of) correlation with the mass ratio should rely on an estimate of the semi-major axis, and not on the measured separation at one specific epoch. The available measurements have been obtained at any orbital phase, and in the case of a significantly eccentric orbit, the measured separation is not necessarily a good proxy for the semi-major axis. 

\section{Discussion}\label{sec:discussion}

\subsection{Outer working angle}\label{disc:OWA}

As we limited our search for companions to OWA, the results for the companions found close to this limit (separation > 40 mas, HD~47839, HD~193322, HD~202214, HD~207198, and HD~206267) might correspond to a local minimum of $\chi^{2}$, while the global minimum may not be probed in our search. 
The detection of those companions is still valid, as the interferometric signal still favors the presence of a companion compared to a uniform disk, but the found separation should be considered as a lower limit. 
Further observations with a wider OWA could confirm the real separation of those companions, but this requires a higher spectral resolution, which is doable with MIRC-X, however, it would reduce the sensitivity of the instrument. It can be achieved only for the brightest systems. Alternative methods such as sparse aperture masking may thus be preferred.

\subsection{Candidates for orbital parameter measurements}

While the statistical analysis will require the large survey data, the currently detected companions can already give us good candidates for follow-up orbital parameter measurements. To be a good candidate for follow-up with interferometry, we take here a limit on the orbital period of a maximum of 10 years, which is a reasonable time scale for follow-up observations. 

To obtain an estimation of the orbital period of the detected companions, we use the Kepler's third law:
\begin{equation}
    P^2 = \frac{a^3}{M_1+M_2}
\end{equation}
where $P$ is the period in years, $a$ is the semi-major axis of the orbital ellipse in AU, and $M_1$ and $M_2$ are the masses of the two objects orbiting each other, in solar mass.

For our estimation of $P$, we approximate the semi-major axis $a$ with the projected separation that we computed earlier. This will lead to an overestimate of $a$, hence $P$, for eccentric systems as companions spend a larger fraction of the orbit at separations $> a$ (Kepler's law). For the mass of the two objects, we use the $m_{\text{evol}}$ in table~4 of \citet{2010A&A...524A..98W}, which gives a theoretical mass of O-type stars as a function of their spectral type. We used the spectral types of our objects listed in Table~\ref{tab:logObs} to get an estimation of the mass for the central object, and we used the estimated mass ratio $q$ for the detected companion mass.

With the flux ratio, separation, and estimated mass ratio, we could estimate also the shift of the photo-center of our binary systems. This estimated shift could give us candidates of systems for which we could determine the orbital parameters with Gaia, as discussed in~\citet{2017A&A...601A..34L}. From equation five in that paper we can determine that:
\begin{equation}
    \mu = \frac{a(q-f)}{(f+1)(q+1)}
\end{equation}

For our estimation of $\mu$, we used the separation of our detected companions as an estimation of $a$, the estimated mass ratio $q$, and the detected flux ratio $f$.

\begin{figure}[!ht]
\centering
\includegraphics[width=\columnwidth]{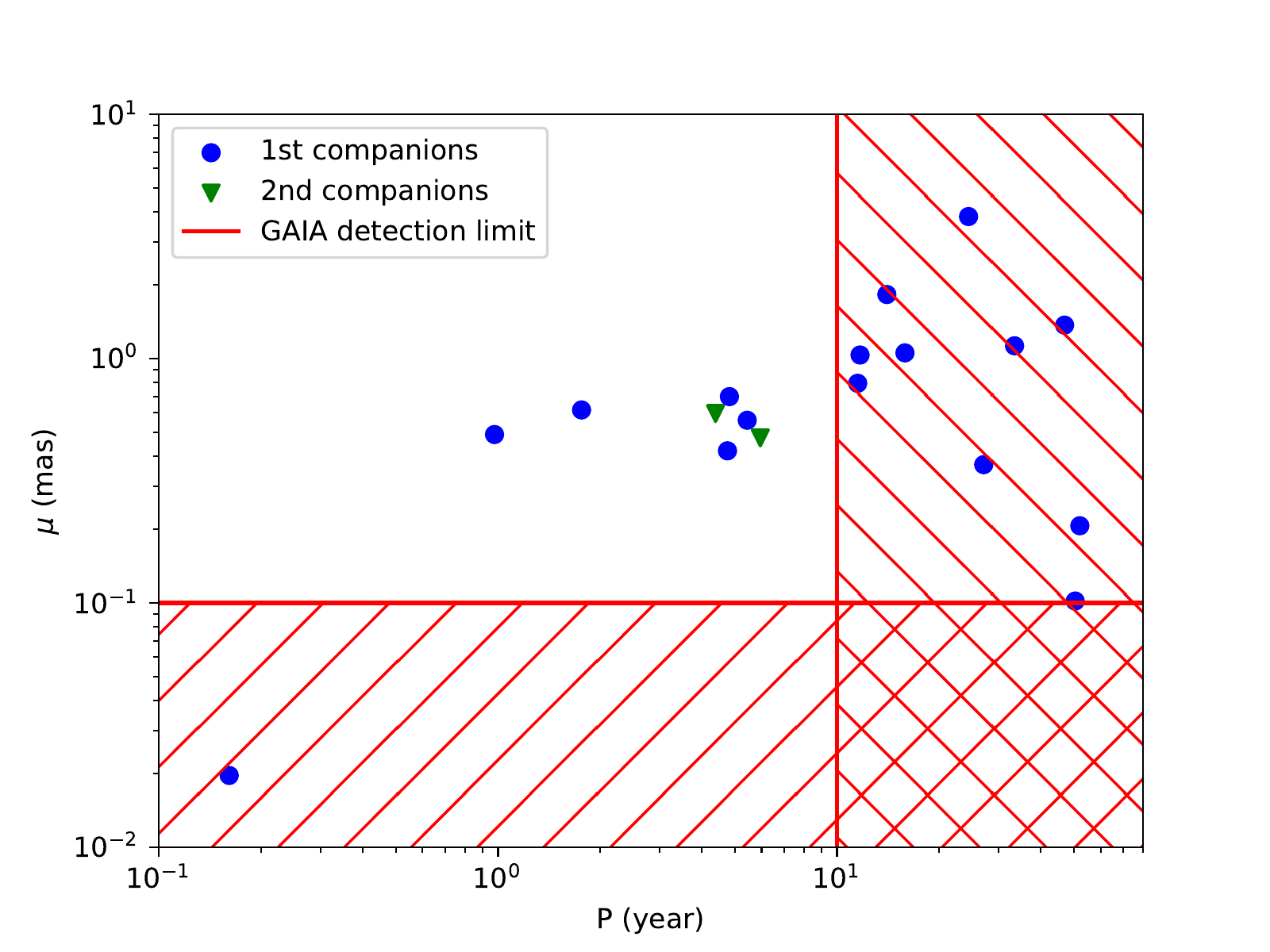}
\caption{Estimated photo-center shift $\mu$ in mas as a function of the estimated orbital period in years. The red lines show the limit of capability of Gaia to measure the orbital parameters of a system, with the hatched area being not accessible by Gaia.} \label{fig:P_mu}
\end{figure}
Figure~\ref{fig:P_mu} displays the estimated photo-center shift $\mu$ in mas as a function of the estimated orbital period in years. The detection limit on $\mu$ detectable by Gaia is set to 0.1~mas or higher. This value comes from the fact that Gaia needs a shift of at least three times the Gaia accuracy of a transit, which is 0.034~mas on bright targets ($G < 12$, see Sect.~4 of \citet{2014ApJ...797...14P}), which is the case for all our targets. The conservative limit in the period for which Gaia can determine the orbital parameters is set to 6 years. This limit is the one for which Gaia will be able to determine orbital parameters from a detectable astrometric shift of the photo-center in almost any case. But this limit is for the nominal mission period of 5 years. Now Gaia has observed for 7.8 years and the anticipated mission lifetime is now at least 10 years. So we estimate the expected limit on Fig.~\ref{fig:P_mu} at 10 years, which corresponds also to the period we set for the interferometric survey.

First, we see that eight companions have an estimated orbital period of less than 10 years, which makes them suitable to be followed up by interferometry directly.
Then, we see that Gaia should be able to obtain the orbital parameters of seven of our systems, which have an estimated orbital period of approximately 6 years or less and an estimated photo-center shift of more than 0.1~mas. For the periods larger than 10 years, the orbital parameters will not be well constrained by Gaia, but combined with the data from interferometry, it should be possible to constrain orbital parameters well.
Furthermore, the orbital parameters with Gaia are interesting because, in combination with the information from interferometry, one could measure the individual masses of each component of a multiple system. Currently interferometry needs to be combined with spectroscopy to measure individual masses (e.g.,~\citealt{2017A&A...601A..34L,2018A&A...616A..75M,2021A&A...651A.119F,2022arXiv220413075S}). The technique with Gaia would not need spectroscopy because it will provide the distance, which with the orbit size from interferometry would provide the total mass of the system, and by combining the orbit size and the orbit size of the photo-center provided by Gaia, one would obtain the mass ratio of the components of the system. Combining both total mass and mass ratio would provide the individual masses of these components. Gaia would also probe a different range of separation than the spectroscopy. These two techniques would then be complementary. In addition, one would be able to compare the results of the two methods and remove the potential biases that each of these methods could present, as the range of masses estimated by spectroscopy alone shows discrepancies with the measured masses by combining spectroscopy and interferometry~\citep{2017A&A...601A..34L}.

\section{Conclusion}\label{sec:conclusion}

From the results of this pilot study, we can conclude that a large survey to study the multiplicity of northern O-type stars can be performed with the instrument MIRC-X at the CHARA Array. Indeed, we demonstrated that we can constrain the multiplicity of O-star systems with a magnitude in the H-band around 7.5~mag, in good atmospheric conditions. This magnitude allows the observation of more than 120 northern O-type stars. From the experience gained with this pilot survey, we can observe six to eight science targets per night with normal conditions. Taking into account an average loss of 25\% of the night due to bad atmospheric conditions weather or technical issues, we estimate that the large program will require approximately 25 nights to be completed.

This study also detected 19 companions in 17 different systems, including 13 companions detected for the first time, notably the companion responsible for the nonthermal emission in Cyg OB2-5 A, and the confirmation of the candidate companion of HD~47129 previously suggested by SMASH+. The preliminary statistical study gives us a multiplicity fraction $f_{\text{m}} = 17 / 29 = 0.59 \pm 0.09$, and a companion fraction of $f_{\text{c}} = 19 / 29 = 0.66 \pm 0.13$. Those results are consistent with the results of the southern large survey already performed, SMASH+.

We also demonstrated that a number of the detected systems are suitable for follow-up studies for orbital parameter measurement, either by interferometry (8) and/or with Gaia (7). The results obtained in this study are promising in terms of scientific returns of a more ambitious project focusing on a large sample, and involving repeated observations spread over several years. 

\begin{acknowledgements} 
We would like to thank Carine Babusiaux, from IPAG, for her help in determining the capability of Gaia to help in the constraint of orbital parameters and Laetitia Rodet in her explanation of the relevance of the study of the Kozai-Lidov cycle. We also thank the referee for their constructive comments that resulted in a better paper.
L.M. and E.G. acknowledge the European Space Agency (ESA) and the Belgium Federal Science Policy Office (BELSPO) for their support in the framework of the prodex programme.
This work is based upon observations obtained with the Georgia State University Center for High Angular Resolution Astronomy Array at Mount Wilson Observatory. The CHARA Array is supported by the National Science Foundation under Grant No. AST-1636624 and AST-2034336. Institutional support has been provided from the GSU College of Arts and Sciences and the GSU Office of the Vice President for Research and Economic Development.
Time at the CHARA Array was granted through the NOIRLab community access program (NOIRLab 2018A-0158, 2018B-0123, 2019A-0080; PI: C. Lanthermann).
S.K., N.A., and C.L.D. acknowledge support from an ERC Starting Grant (Grant Agreement No.\ 639889), ERC Consolidator Grant (Grant Agreement ID 101003096), and STFC Consolidated Grant (ST/V000721/1). A.L. received funding from STFC studentship No.\ 630008203.
J.D.M. acknowledges funding for the development of MIRC-X (NASA-XRP NNX16AD43G, NSF-AST 1909165) and MYSTIC (NSF-ATI 1506540, NSF-AST 1909165).
This project has received funding from the European Research Council (ERC) under the European Union’s Horizon 2020 research and innovation programme (Grant agreement No. DLV-772225: MULTIPLES).
This research has made use of the Jean-Marie Mariotti Center \texttt{Aspro} service \footnote{Available at \href{http://www.jmmc.fr/aspro}{http://www.jmmc.fr/aspro}}. This work was supported by the Programme National de Physique Stellaire (PNPS) of CNRS/INSU co-funded by CEA and CNES. This work has been partially supported by the LabEx FOCUS ANR-11-LABX-0013. This research has made use of the SIMBAD database, operated at CDS, Strasbourg, France.

\end{acknowledgements}

\bibliographystyle{aa}
\bibliography{mybib}

\begin{appendix}

\section{Calibrators}\label{Ann:cal}

In this section, we list the stars that we used as calibrators in data reduction in Table~\ref{tab:cal-used}. In addition, we observed eight targets to use them as calibrators but ended up showing resolved features such as nonconstant closure phases, or obvious multiplicity signal. These targets are 2MASS~J21472749+5746367, HD~47415, HD~167882, HD~194479, HD~237112, HD~239737, HD~281311, and SAO~49700: they will be reported and added into the bad calibrator list\footnote{\href{https://www.jmmc.fr/badcal/}{https://www.jmmc.fr/badcal/}} of the JMMC.

\longtab[1]{
\begin{longtable}{ccccc}
\caption{\label{tab:cal-used} Calibrator observed and used to reduce the data.
}\\
\hline\hline
Calibrator's name & Diameter [mas] & Error [mas] & Science target & Observing dates \\\hline
\endfirsthead
\caption{continued.}\\
\hline\hline
Calibrator's name & Diameter [mas] & Error [mas] & Science target & Observing date \\\hline
\endhead
\hline
\endfoot
2MASS J19524977+4836263 & 0.46  & 0.03 & HD 188209 & 2018-06-09\\
2MASS J20295613+4137570 & 0.31  & 0.01 & Cyg OB2-9 & 2018-09-23\\
... & ...  & ... & HD 193322 &  2018-06-12\\
... & ...  & ... & HD 34078 & 2017-12-13\\
2MASS J20473119+3629045 & 0.2   & 0.01 & HD 201345 & 2019-06-27 \\
2MASS J21281481+6005284 & 0.52  & 0.03  & HD 209975 & 2018-06-09 \\
HD~21820    & 0.48  & 0.03  & HD 24534 & 2018-09-23, 2018-09-27\\
HD~22269    & 0.58  & 0.03  & HD 24534 & 2018-09-27\\
HD~24688    & 0.37  & 0.03  & HD 24431 & 2018-09-24\\
HD~26311    & 1.64  & 0.03  & HD 24534 & 2018-02-08\\
HD~30111    & 0.562 & 0.03  & HD 34078 & 2017-12-13\\
HD~30793    & 0.53  & 0.03 & HD 34078 & 2018-09-23, 2018-09-27\\
HD~32518    & 0.816 & 0.03  & HD 30614 &  2018-09-25\\
HD~35238    & 1.03  & 0.05  & HD 34078 & 2018-09-27\\
HD~42618    & 0.377 & 0.03  & HD 34078 & 2017-12-13\\
HD~45089    & 0.61  & 0.03  & HD 41129 & 2018-09-23, 2018-09-24 \\
...   & ...  & ...  & HD 45314 & 2018-09-25 \\
HD~46714    & 0.86  & 0.03  & HD 47432 & 2018-09-27 \\
HD~48596    & 0.53  & 0.03  & HD 47432 & 2018-09-26, 2018-09-27\\
HD~49019    & 0.89  & 0.03  & HD 36861 & 2018-02-07 \\
HD~52961    & 0.232 & 0.02  & HD 34078 & 2017-12-13 \\
HD~77250    & 0.8   & 0.04  & HD 34078 & 2017-12-13\\
HD~87828    & 0.78  & 0.03  & HD 89353 & 2018-02-07 \\
HD~150470   & 0.45  & 0.01  & HD 150193$^{\ast}$ & 2019-06-24 \\
HD~155524   & 0.49  & 0.03 & Cyg OB2-5 A & 2018-06-10\\
...   & ...  & ... & HD 157214$^{\ast}$ &  2018-06-11\\
HD~165480   & 0.44  & 0.01  & HD 166737 & 2019-06-24\\
HD~185663   & 0.64  & 0.03  & HD 188001 & 2018-09-26 \\
HD~186962   & 0.41  & 0.01  & HD 188001 & 2018-09-26\\
HD~189942   & 1.15  & 0.05  & HD 193237$^{\ast}$ & 2018-06-24\\
HD~193217   & 1.57  & 0.03  & Cyg OB2-5 A & 2018-06-10\\
HD~195647   & 0.42  & 0.03& HD 229196&  2018-09-24 \\
...   & ...  & ...&  Cyg OB2-9 &   2018-09-27 \\
...  & ...  & ...&  Cyg OB2-8 &  2018-06-12 \\
...  & ...  & ...& Cyg OB2-10 &   2019-06-23 \\
HD~196134   & 0.71  & 0.06 & Cyg OB2-5 A &  2019-06-23 \\
HD~196360   & 0.6   & 0.03& Cyg OB2-9 & 2018-09-24  \\
...   & ...   &...& HD 229196 & 2018-09-27  \\
HD~200060   & 0.516 & 0.03 & Cyg OB2-11 & 2018-09-25\\
HD~204050   & 0.42  & 0.03 & HD 203064 & 2018-06-10\\
HD~204721   & 0.6   & 0.06  & HD 207198 & 2019-06-24\\
HD~211982   & 0.57  & 0.02  & HD 202214 & 2018-06-09 \\
...   & ...  & ...  & HD 210839 & 2018-06-09 \\
HD~212289   & 0.48  & 0.03  & HD 203064 & 2018-06-10 \\
...   & ...  & ...  & HD 214686 & 2018-06-11 \\
HD~217711   & 0.67  & 0.03  & HD 2018086 & 2018-09-26\\
HD~228660   & 0.58  & 0.01  & HD 228779 & 2019-06-26\\
HD~228721   & 0.31  & 0.01  & HD 229202 & 2019-06-27 \\
HD~228852   & 0.54  & 0.03  & Cyg OB2-9 & 2018-09-24 \\
...   & ...  & ...  &  HD 193322 &  2018-09-27, 2019-06-24\\
HD~232948   & 0.68  & 0.03  & HD 36861 & 2018-02-08 \\
...   & ...  & ...  & HD 28446 & 2018-09-24 \\
HD~235757   & 0.43  & 0.01  & HD 210809 & 2019-06-26 \\
HD~235872   & 0.34  & 0.01   & HD 218915 & 2019-06-27\\
HD~237032   & 0.76  & 0.03  & HD 17505 & 2018-09-25, 2018-09-26\\
HD~237036   & 0.5   & 0.03  & HD 17505 & 2018-09-26\\
HD~239636   & 0.36  & 0.01  & HD 202214 & 2018-06-09, 2019-06-24\\
HD~239760   & 0.61  & 0.02  & HD 206267 & 2019-06-22 \\
HD~240211   & 0.34  & 0.03  & HD 217086 & 2018-09-26 \\
HD~246454   & 0.6   & 0.03  & HD 36861 & 2018-09-26\\
HD~254874   & 0.51  & 0.03  & HD 45314 & 2018-02-07\\
HD~261683   & 0.56  & 0.03  & HD 47839 & 2018-09-25\\
SAO 50079   & 0.4   & 0.03  & Cyg OB2-10 & 2018-06-12\\
SAO 50138   & 0.38  & 0.03  & HD 199579 & 2018-06-11\\
SAO 50258   & 0.27  & 0.03  &  LS III +46 11 & 2018-06-24 \\
\end{longtable}
\tablefoot{The first column is the calibrator name. The second column is the diameter of the calibrator in the H-band in mas as stated in the JSDC catalog \citep{2017yCat.2346....0B}. The third column is the associated error in mas. The fourth column is the science target it has calibrated. The fifth column gives the observing dates at which the calibrator was used for this science target. We note that science targets noted with the symbol $\ast$ are non O-type stars observed during the observing runs: their calibrators were still used for the data reduction. Concerning O-type star science targets present in this table that were observed with the associated calibrator, but had not good enough quality data, thus them not being present in the results, their calibrator was still used for the data reduction.}

}

\section{Coherent time used in the data reduction}

In this section, we display the number of frames that we coherently add to optimize the quality of the data in Table~\ref{tab:cohtime}.

\begin{table}[ht!]
\caption{Number of frames coherently added for V2 and CP for each observation.}
\label{tab:cohtime}
\centering
\begin{tabular}{cccc}\hline\hline
Target's name & DATE-OBS  & V2  & CP \\
&(YYYY-MM-DD)&(n coadd)&(n coadd)\\\hline
Cyg OB2-5 A & 2018-06-10 & 8 & 11  \\ 
... & 2019-06-23 & 8 & 9 \\ 
Cyg OB2-9 & 2018-09-27 & 11 & 3 \\ 
Cyg OB2-10 & 2018-06-12 & 15 & 11  \\ 
HD~17505 & 2018-09-26 & 9 & 3 \\ 
HD~19820 & 2018-09-26 & 9 & 3 \\ 
HD~24431 & 2018-09-24 & 3 & 3 \\ 
HD~28446 & 2018-02-08 & 12 & 9 \\ 
... & 2018-09-24 & 3 & 3 \\ 
HD~30614 & 2018-09-25 & 11 & 3 \\ 
HD~34078 & 2017-12-13 & 10 & 10 \\ 
... & 2018-09-27 & 11 & 3 \\ 
HD~36861 & 2018-02-07 & 8 & 5 \\ 
... & 2018-09-26 & 9 & 3 \\ 
HD~45314 & 2018-09-25 & 11 & 3 \\ 
HD~47129 & 2018-09-24 & 3 & 3 \\ 
HD~47432 & 2018-09-27 & 11 & 3 \\ 
HD~47839 & 2018-09-25 & 11 & 3 \\ 
HD~167971 & 2019-06-24 & 9 & 5 \\ 
HD~188001 & 2018-09-26 & 9 & 3 \\ 
HD~193322 & 2018-06-12 & 15 & 10 \\ 
... & 2019-06-24 & 9 & 5 \\ 
HD~195592 & 2018-06-11 & 15 & 15 \\ 
HD~201345 & 2019-06-27 & 8 & 9 \\ 
HD~202214 & 2019-06-24 & 10 & 15 \\ 
HD~206183 & 2019-06-27 & 8 & 9 \\ 
HD~206267 & 2019-06-26 & 8 & 15 \\ 
HD~207198 & 2019-06-24 & 9 & 5 \\ 
HD~209975 & 2018-06-09 & 15 & 10 \\ 
HD~210809 & 2019-06-26 & 8 & 8 \\ 
HD~210839 & 2019-06-23 & 8 & 11 \\ 
HD~217086 & 2018-09-26 & 9 & 3 \\ 
HD~228779 & 2019-06-26 & 8 & 9 \\ 
HD~229196 & 2018-09-27 & 11 & 3 \\ 
\hline
\end{tabular}
\tablefoot{The first column gives the identifier of the system. The second column gives the UT date the data have been taken. The third and fourth columns are the number of frames coherently added for respectively V2 and CP.}
\end{table}

\begin{figure}[!ht]
\centering
\includegraphics[width=\columnwidth]{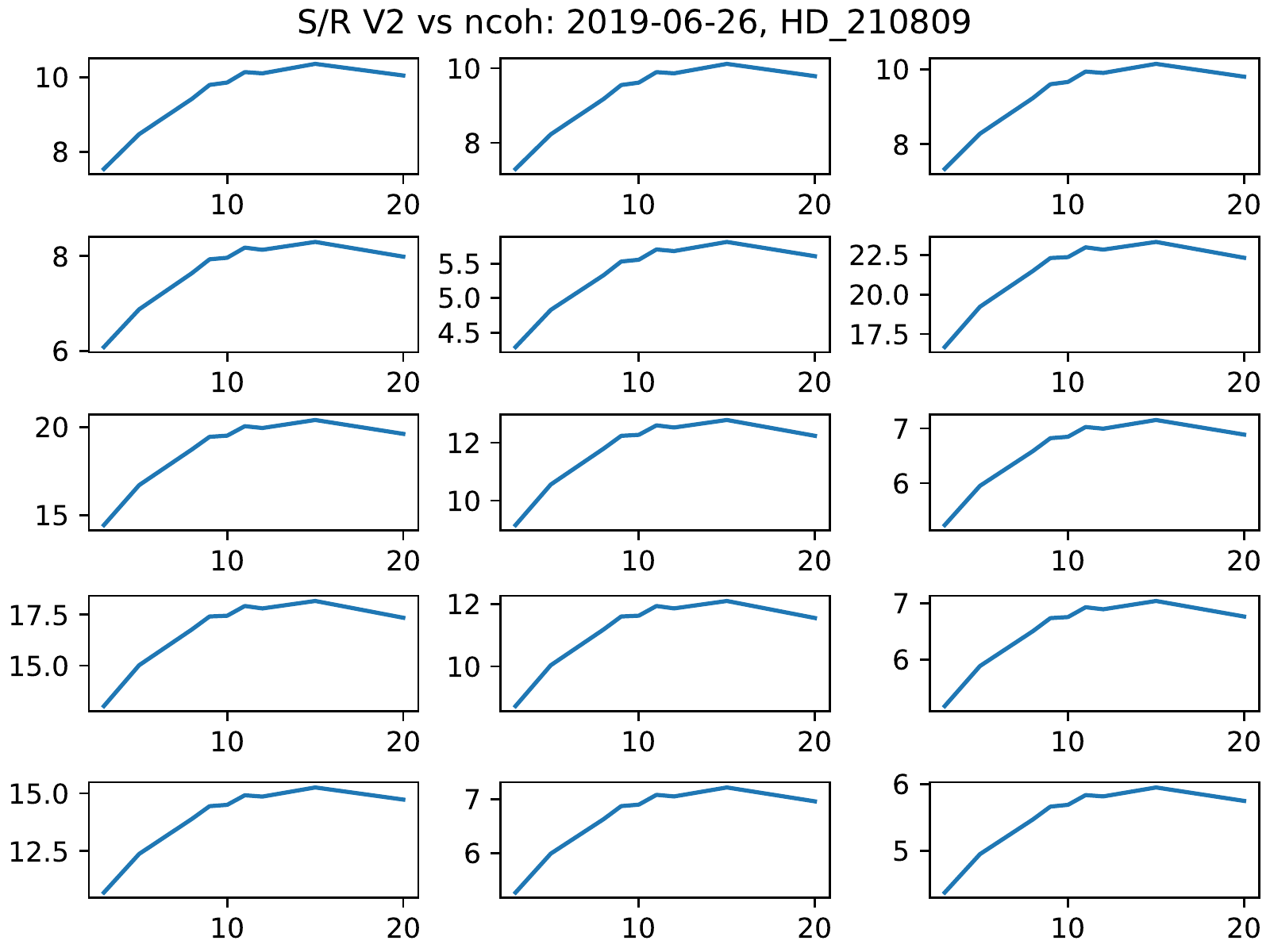}
\caption{V2 S/N as a function of the number of frames added coherently, for HD~210809, for the different baselines.} \label{fig:V2-coadd}
\end{figure}

In Figure~\ref{fig:V2-coadd}, we show an example of S/N of V2 as a function of the number of frames added coherently, for HD~210809, for the date 2019-06-26.

\begin{figure}[!ht]
\centering
\includegraphics[width=\columnwidth]{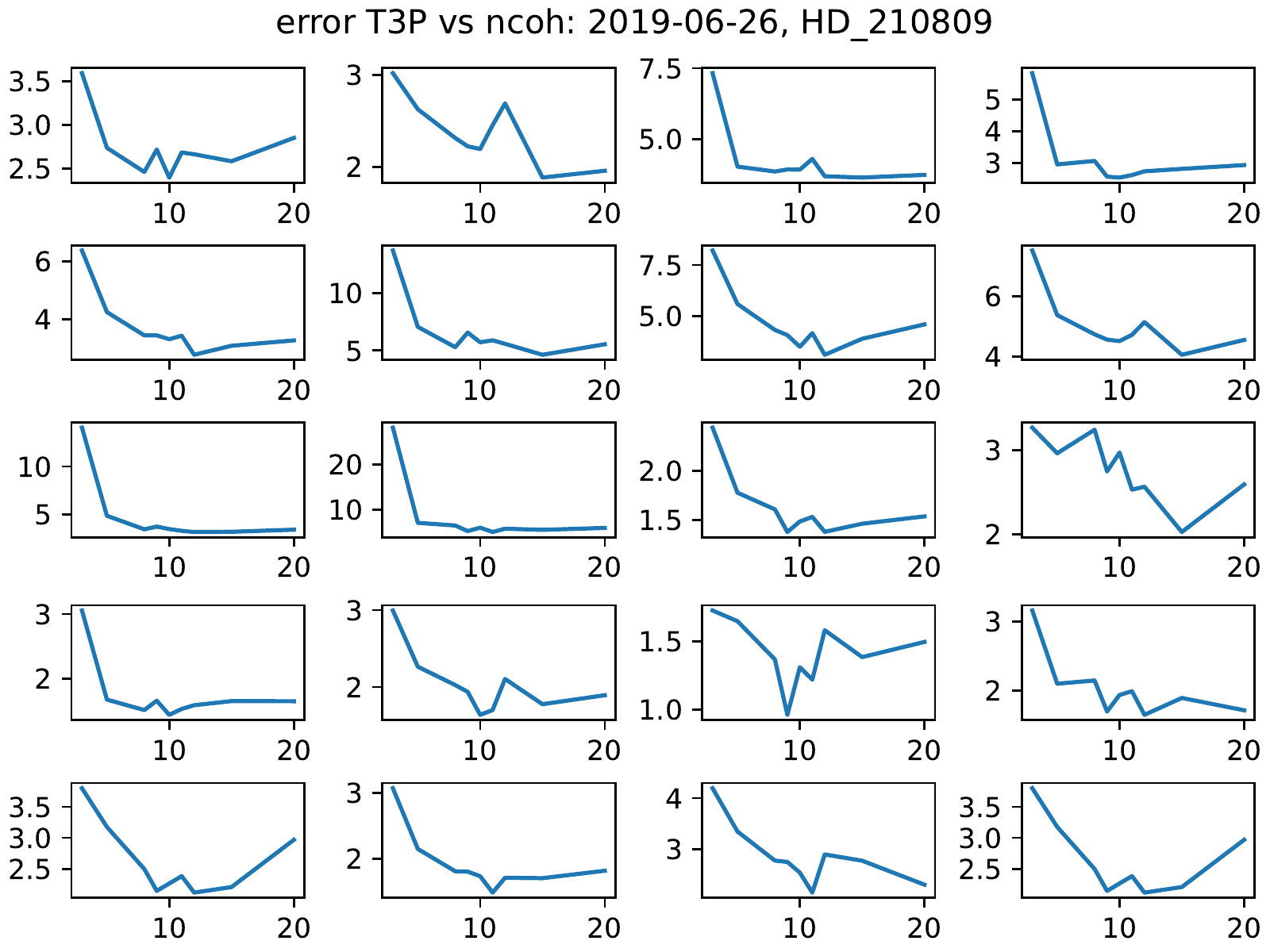}
\caption{CP error as a function of the number of frames added coherently, for HD~210809, for the triplets of telescopes.} \label{fig:T3P-coadd}
\end{figure}

In Figure~\ref{fig:T3P-coadd}, we show an example of the error on the CP as a function of the number of frames added coherently, for HD~210809, for the date 2019-06-26.

\end{appendix}

\end{document}